\documentclass[twocolumn]{aastex631}

\usepackage{amsmath}
\usepackage{graphicx}
\usepackage{txfonts}
\usepackage{multirow}
\usepackage{amssymb}
\usepackage{units}
\usepackage{float}
\usepackage{natbib}
\bibpunct{(}{)}{;}{a}{}{,}
\usepackage{color, colortbl}
\usepackage{xspace}
\usepackage{url}
\usepackage{arydshln}
\usepackage[all]{hypcap}
\usepackage{enumitem,booktabs}
\usepackage[referable]{threeparttablex}
\renewlist{tablenotes}{enumerate}{1}

\newcommand{\xmm}{XMM{\it-Newton}\xspace}
\newcommand{\chandra}{{\it Chandra}\xspace}

\newcommand{\asca}{{\it ASCA}\xspace}
\newcommand{\bepposax}{{\it BeppoSAX}\xspace}

\newcommand{\rosat}{{\it ROSAT}\xspace}

\newcommand{\arcus}{{\it Arcus}\xspace}
\newcommand{\athena}{{\it Athena}\xspace}

\newcommand{\kms}{\ensuremath{\mathrm{km\ s^{-1}}}\xspace}

\newcommand{\ang}{\AA\xspace}

\newcommand{\ovii}{\ion{O}{vii}\xspace}
\newcommand{\oviii}{\ion{O}{viii}\xspace}
\newcommand{\oix}{\ion{O}{ix}\xspace}

\newcommand{\fexix}{\ion{Fe}{xix}\xspace}
\newcommand{\fexx}{\ion{Fe}{xx}\xspace}

\newcommand{\fexxv}{\ion{Fe}{xxv}\xspace}
\newcommand{\fexxvi}{\ion{Fe}{xxvi}\xspace}

\newcommand{\NH}{\ensuremath{N_{\mathrm{H}}}\xspace}
\newcommand{\nh}{\ensuremath{n_{\mathrm{H}}}\xspace}

\newcommand{\teq}{{\ensuremath{t_{\rm eq}}}\xspace}
\newcommand{\tvar}{{\ensuremath{t_{\rm var}}}\xspace}
\newcommand{\trec}{{\ensuremath{t_{\rm rec}}}\xspace}
\newcommand{\lion}{{\ensuremath{L_{\rm ion}}}\xspace}
\newcommand{\spex}{{\textsc{Spex}}\xspace}
\newcommand{\pion}{\texttt{pion}\xspace}
\newcommand{\tpho}{\texttt{tpho}\xspace}

\definecolor{Gray}{gray}{0.95}
\definecolor{Cyan}{rgb}{0.9,1,1}
\definecolor{Red}{rgb}{0.992, 0.8, 0.69}
\DeclareRobustCommand{\ion}[2]{\textup{#1\,\textsc{\lowercase{#2}}}}

\shorttitle{TPHO: Time-dependent photoionisation model}
\shortauthors{Daniele Rogantini, Missagh Mehdipour, et al.}
\graphicspath{{./}{plot/}}

\begin{document}

\title{TPHO: a time-dependent photoionisation model for AGN outflows\footnote{Accepted for publication on ApJ on 07-Oct-2022}.}

\author[0000-0002-5359-9497]{Daniele Rogantini}
\affiliation{MIT Kavli Institute for Astrophysics and Space Research,
              Massachusetts Institute of Technology,
              Cambridge, MA 02139, USA}
\affiliation{SRON Netherlands Institute for Space Research,
			 Niels Bohrweg 4, 2333 CA Leiden, The Netherlands}             

\author{Missagh Mehdipour}
\affiliation{Space Telescope Science Institute, 3700 San Martin Drive,
             Baltimore, MD 21218, USA}

\author{Jelle Kaastra}
\affiliation{SRON Netherlands Institute for Space Research,
			 Niels Bohrweg 4, 2333 CA Leiden, The Netherlands}
\affiliation{Leiden Observatory, Leiden University,
			 PO Box 9513, 2300 RA Leiden, The Netherlands} 

\author{Elisa Costantini}
\affiliation{SRON Netherlands Institute for Space Research,
			 Niels Bohrweg 4, 2333 CA Leiden, The Netherlands}
\affiliation{Anton Pannekoek Astronomical Institute,
			 University of Amsterdam, P.O. Box 94249,
			 1090 GE Amsterdam, The Netherlands}

\author{Anna Jur\'a\v{n}ov\'a}
\affiliation{SRON Netherlands Institute for Space Research,
			 Niels Bohrweg 4, 2333 CA Leiden, The Netherlands}
\affiliation{Anton Pannekoek Astronomical Institute,
			 University of Amsterdam, P.O. Box 94249,
			 1090 GE Amsterdam, The Netherlands} 

\author{Erin Kara}
\affiliation{MIT Kavli Institute for Astrophysics and Space Research,
              Massachusetts Institute of Technology,
              Cambridge, MA 02139, USA}



\begin{abstract}

Outflows in active galactic nuclei (AGN) are considered a promising candidate for driving AGN feedback at large scales. However, without information on the density of these outflows, we cannot determine how much kinetic power they are imparting to the surrounding medium. Monitoring the response of the ionisation state of the absorbing outflows to changes in the ionising continuum provides the recombination timescale of the outflow, which is a function of the electron density. We have developed a new self-consistent time-dependent photoionisation model, \tpho, enabling the measurement of the plasma density through time-resolved X-ray spectroscopy. The algorithm solves the full time-dependent energy and ionisation balance equations in a self consistent fashion for all the ionic species. The model can therefore reproduce the time-dependent absorption spectrum of ionized outflows responding to changes in the ionizing radiation of the AGN. We find that when the ionised gas is in a non-equilibrium state its transmitted spectra are not accurately reproduced by standard photoionisation models. Our simulations with the current X-ray grating observations show that the spectral features identified as a multiple-components warm absorbers, might be in fact features of a time-changing warm absorber and not distinctive components. The \tpho model facilitates accurate photoionisation modelling in the presence of a variable ionising source, thus providing constraints on the density and in turn the location of the AGN outflows. Ascertaining these two parameters will provide important insight into the role and impact of ionised outflows in AGN feedback.

\end{abstract}



\section{Introduction}
\label{sec:intro}
Accreting super-massive black holes (with mass between $10^{6} - 10^{9}\ \rm M_{\odot}$) are the central engines of active galactic nuclei (AGN). About half of Seyfert I galaxies show outflows of material in form of ultra-fast outflows, warm absorbers, and ionised/neutral/molecular outflows \citep[e.g.,][]{Costantini07,Fabian12,Tombesi13,Veilleux13,Kaastra14,Tombesi15,Fiore17,Laha21}. In the last decades, the potential importance of AGN outflows for the growth of super-massive black holes, the enrichment of the intergalactic medium, the evolution of the host galaxy, cluster cooling flows and the luminosity function of AGN has been widely recognised \citep{Fabian12,Gaspari20}. 

X-ray spectroscopy is a powerful tool to study ionised outflows closer to the central region. Measurement of absorption line spectra yield reliable information on different aspects of the outflows such as their kinematics (turbulence and outflow velocity) and ionisation state. Spectral studies of warm absorbers highlight their complex multiphase structure, which spans a wide range of ionisation parameters, $\log({\xi / \rm erg\ cm\ s^{-1}}) = -1 \rm \ to\ 3$, and column densities, $\NH = 10^{21-23}\ \rm cm^{-2}$ \citep{Blustin05,McKernan07,Krongold09,Tombesi13,Laha14,Behar17}. They are usually detected as absorption lines and edges from H-like and He-like ions of the most abundant elements, such as C, O, N, Ne, Mg, Si, Si and Fe, in the soft spectra below 2~keV \citep{Crenshaw03,Mehdipour10,Mehdipour18,Ebrero21}. Low ionisation absorbers ($\log \xi=0-1.5$) imprint deep spectral absorption troughs in the rest frame wavelength $15-17$~\ang due to the blended Fe M shell unresolved transitions array \citep[UTA;][]{Behar01}. All these spectral features are always found blueshifted with respect to the systematic redshift, implying that these absorbers are outflowing with a velocity, $v_{\rm out}$, between $\sim100$ and $\sim5000$~\kms \citep{Laha21}. 

Spectroscopy on its own does not provide information regarding the distance $r$ of the absorber material to the central ionising source. This can be seen within the definition of the ionisation parameter $\xi = L_{\rm ion}/(n_{\rm e}r^2)$ where \lion is the ionising luminosity of the source facing the cloud (integrated between $1-1000$ Ryd) and $n_{\rm e}$ is the electron density of the gas \citep{Tarter69,Krolik81}. With \lion and $\xi$ known from observations, only the product $n_{\rm e}r^2$ can be determined. This degeneracy makes it challenging to assess the significance of ionised outflows in AGN feedback. The energetics of a spherical shell-like outflow can be quantified by the mass outflow rate, $\dot{M}_{\rm out} \simeq 1.23m_{\rm p}\NH r v_{\rm out}\Omega$ and kinetic luminosity, $\dot{E}_{\rm k}=\frac{1}{2}\dot{M}_{\rm out}v_{\rm out}^2$, where the constant 1.23 takes into account the abundances of elements, $m_{\rm p}$ is the proton mass, and $\Omega$ is the solid angle subtended by the outflow, which is expected to be $\sim \pi$ \citep{Blustin05}. Determining these two quantities requires precise measurements of the outflow distance to the central source.

A common approach to characterising the outflow location is measuring the density $n_{\rm e}$ of the plasma and deriving the distance through the definition of the ionisation parameter. One way to constrain the density via spectroscopy is to use density-sensitive absorption lines from metastable levels \citep{Kraemer06,Arav15}. This method is widely used in UV spectroscopy, where these metastable transitions of \ion{C}{ii}*, \ion{C}{iii}*, \ion{S}{iii}* and \ion{Fe}{ii}* are commonly detected. In the X-ray band, the density diagnostic of AGN outflows using absorption lines is not very effective. Low signal to noise ratio of these rather weak X-ray lines makes their density diagnostic inaccessible in most AGN \citep{Kaastra04,King12,Mao17}. Only the enhanced sensitivity of future X-ray telescopes, such as \arcus \citep{Smith16} and \athena \citep{Nandra13}, and their broader wavelength range will enable the detection of the density sensitive lines in the X-ray band opening the doors for a new kind of diagnostic in AGN outflows \citep{Kaastra17}. 

An alternative method to estimate the gas density is a spectral analysis of density sensitive emission lines. The ratio between the recombination, intercombination and forbidden emission lines in the He-like triplets varies as a function of the plasma density \citep{Porquet10}. In the literature there are several studies of the He-like triplets in emission from photoionised plasma in AGN where the upper limits of the density are derived \citep{Porquet00,Collinge01,McKernan03}. However, these spectral analyses are not only limited by the instrumental sensitivity, but also by Li-like absorption lines which can significantly diminish the intensity of the intercombination line in a photoionised medium, leading to more uncertain density diagnostic \citep{Mehdipour15b}. 

The approach that we adopt in this work to determine the gas density is a timing analysis of the response timescale of the plasma to changes in the ionising radiation. How fast the gas responds depends on the recombination timescale, which is a function of the gas density (see Section \ref{sec:time_evo} for a detailed description). Early time-dependent photoionisation studies used only a few spectral features (the most significant) to measure the delay (or the lack of delay) of the ionisation state of the gas relative to the ionising luminosity variation. This constrains the recombination timescale and consequently the density and the distance. 

First attempts of tracing the gas variability with low-resolution CCD instruments, such as \bepposax and \asca were done studying the evolution of the X-ray absorption edges \citep{Morales00}. However, the higher energy resolution of the grating on board \xmm \citep{denHerder01} and \chandra \citep{Canizares05} allowed to meaningfully follow the variation of strong absorption lines. For example, \cite{Behar03} tracked the response of the multi-phase warm absorber in RGS data of NGC~3783 investigating the evolution of \ovii and \oviii lines for the high-ionisation component and the Fe UTA for the low-ionisation component. For the same source, \cite{Reeves04} analyses the variability of the \fexxv line due to the presence of a possible higher ionisation component. 

Subsequently, broad-band time-dependent photoionisation models have been applied to large \xmm and \chandra campaign such as Mrk~509 \citep{Kaastra12}, NGC~5548 \citep{Ebrero16}, and NGC~7469 \citep{Behar17,Mehdipour18,Peretz18}. The complex multi-component outflow in NGC~4051 is one of the best studied cases among Seyfert I galaxies \citep{Silva16}. NGC~4051 is a particularly bright and variable Seyfert I galaxy. \cite{Nicastro99} and \cite{Krongold07} suggested a quick response of the absorber ionisation to the continuum level that places the absorber as close as 0.003 pc. Recently, \cite{Wang22} studied the response of an ionized obscurer in NGC~3227, suggesting a location of  $0.001\ {\rm pc} \lesssim r \lesssim 0.1\  {\rm pc}$ from the central source, whereas the warm absorbers are  further out ($0.01\ {\rm pc} \lesssim r \lesssim 100\  {\rm pc}$).

Here we present a new Time-dependent PHOtoionisation model, \tpho, based on the methodology followed by \cite{Kaastra12} and \cite{Silva16}. We have upgraded the photoionisation code \pion \citep{Mehdipour16} to treat non-photoionisation equilibrium state conditions. The ultimate goal of the model is to derive the density and therefore the location of the AGN outflows through precise high-resolution X-ray spectral-timing analysis. In Section \ref{sec:time_evo}, we describe the time-dependent effects that have been integrated in the new photoionisation model. In Section \ref{sec:tdp}, we show three different applications of the model to describe the impact of time-dependent effects on the inferred plasma physical properties such as temperature, ionic concentration, heating and cooling rates, etc. The importance of the time-dependent effects in high-resolution X-ray spectroscopy studies of AGN outflows is discussed in Section \ref{sec:discussion}. 

\section{Time-evolving ionisation state}
\label{sec:time_evo}
Traditionally, modelling of astronomical photoionised plasma is done assuming the condition of steady-state equilibrium, which means that the gas ionisation is balanced by recombination, atomic excitations are balanced by spontaneous and induced de-excitations, and electron heating is balanced by cooling. In this equilibrium condition, it is possible to calculate the distribution of the ionic species in a cloud of optically thin gas illuminated by an intense ionising luminosity by setting the photoionisation rate equal to the radiative recombination rate \citep[see][]{Netzer90} and adding the condition for charge conservation \citep{Blandford90}. 

The steady equilibrium assumption is valid only if \lion is not varying in time or if the equilibrium timescales for recombination, photoionisation and thermal balance are much shorter than variability timescales in the ionising continuum. 
Whenever the illuminating radiation changes on timescales shorter than the equilibrium timescales, it is necessary to take into account the full time-dependent form of the ionisation balance equations. By definition, the relative density $n_{{X,i}}$ of ion $i$ of a certain element $X$ varies with time as a function of the electron density, $n_{\rm e}$, the recombination rate from stage $i+1$ to $i$ (given by the recombination coefficient $\alpha_{X,\,i+1\rightarrow i}$) times the electron density) and the ionisation rate from stage $i$ to $i+1$. Gathering the contribution by photoionisation, Compton ionisation, collisional ionisation and Auger ionisation in the ionisation coefficient $\gamma_{X,\ i\rightarrow i+1}$, the time-dependent ionisation balance equation is written as follows \citep{Krolik95}:

\begin{multline}
\frac{\textrm{d}n_{X,i}}{\textrm{d}t} =(n_{X,i+1}n_{\rm e}\alpha_{X,i+1\rightarrow i})-(n_{X,i}n_{\rm e}\alpha_{X,i\rightarrow i-1})+ \\
+\sum_{k=1}^{i-1}(n_{X,i-1-k}\gamma_{X,i-1\rightarrow i})-\sum_{k=1}^{i}(n_{X,i-k}\gamma_{X,i\rightarrow i+1}).
\label{eq:diffeq}
\end{multline}

The equation represents the sum of the destruction (second and fourth terms) and formation rate (first and third terms) of each ion considering several processes of ionisation and radiative recombination. The two ionisation terms take into account all forms of ionization in \spex, including the inner-shell ionization that can induce additional multiple ionisations. In case of photoionisation equilibrium, photoionisation dominates over the other ionisation processes including collisional ionisation. 

\begin{figure*}
   \centering
   \includegraphics[width=\hsize]{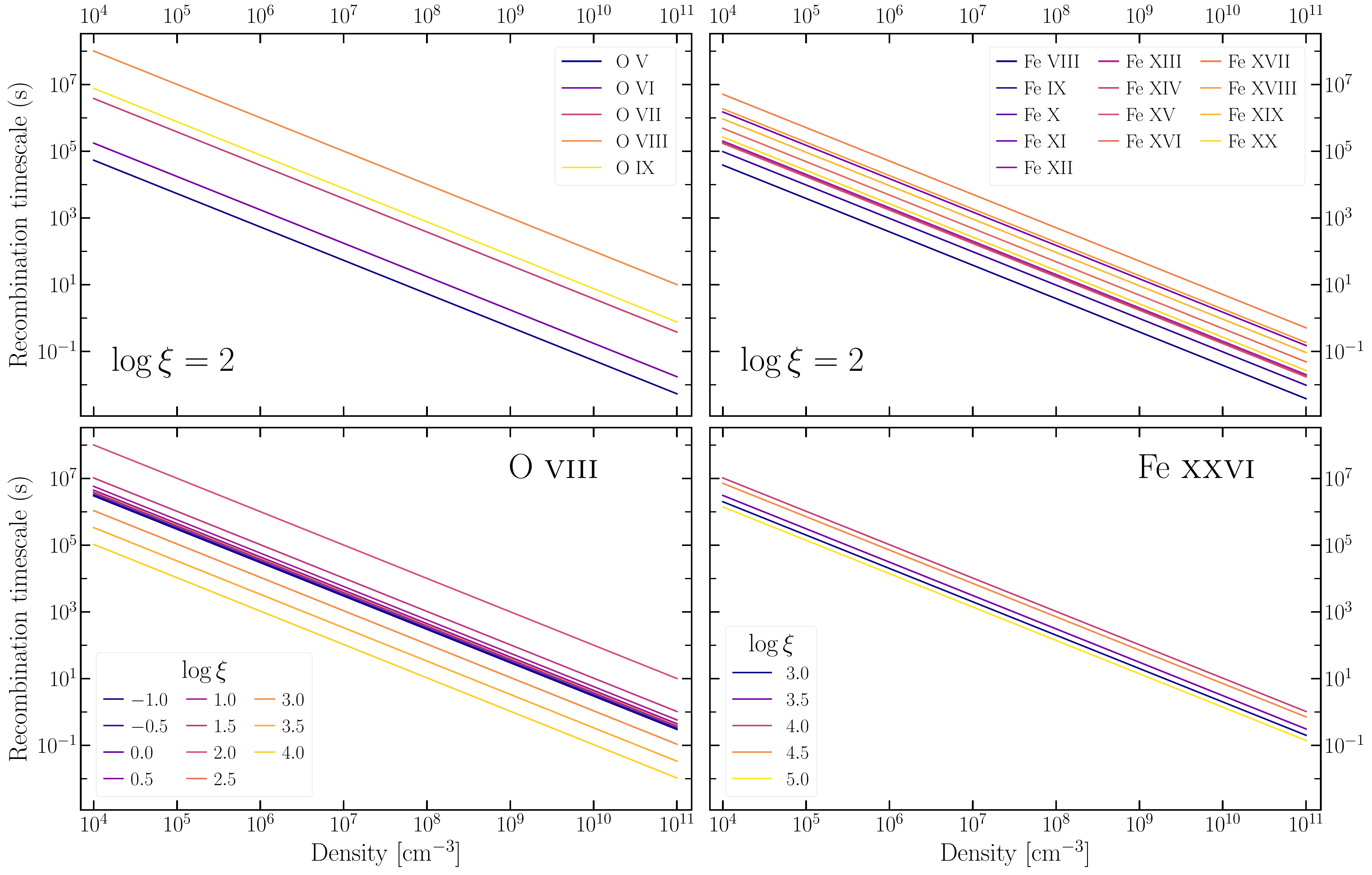}
      \caption{Recombination timescale as a function of the density of a plasma in photoionisation equilibrium. In the top panels we show the recombination timescale of the oxygen (left) and iron (right) ions for a plasma with an ionisation parameter of $\log \xi =2$. We show only the ions which have a relative concentration larger than $10^{-6}$. In the bottom panels, we display the recombination timescales of the H-like ion of oxygen (left) and iron (right) for a grid of ionisation parameters.}
         \label{fig:trec}
   \end{figure*}

The solution of Equation \ref{eq:diffeq} defines the photoionisation timescales, $t_{\rm ph}$, and recombination, $t_{\rm rec}$, timescales which together measure the time necessary for a plasma with density $n_{\rm e}$ to reach photoionisation equilibrium with the ionising continuum for an increasing or decreasing flux phase, respectively. For each point of the light curve of \lion, these timescales can be approximated by the inverse of the destruction rate of the ion $i$ of the element $X$ which define the equilibrium timescale, $t_{\rm eq}$ \citep{Krolik95,Nicastro99}:
\begin{equation}
\teq^{X^i;X^{i+1}} \sim \frac{1}{\alpha_{X,i+1 \rightarrow i}n_{\rm e}} \cdot \bigg[ \frac{1}{-(\alpha_{X,i\rightarrow i-1}/\alpha_{X,i+1 \rightarrow i})+(n_{X,{i+1}}/n_{X,i})} \bigg].
\label{eq:trec}
\end{equation}
Thus the plasma will reach equilibrium with the ionising source after a time delay $t_{\rm eq}$. Equation \ref{fig:trec} shows that the time necessary for a gas to reach equilibrium critically depends on the characteristic electron density $n_{\rm e}$ of the plasma. High-density clouds reach the photoionisation equilibrium in short timescales by quickly responding to changes in the ionising continuum. On the other hand, low-density gases need more time to achieve the ionisation balance with the ionising radiation. Especially in this late response scenario due to low-density absorbers it is important to consider the time-dependent effect in the photoionisation modelling.

The behaviour of the time-variation of ionic abundances depends on two factors: first the ratio $\teq/\tvar$, where \tvar is the typical timescale on which most of the fluctuations in the ionising radiation occurs. Second it depends on the coefficient of variation (i.e. the ratio of the standard deviation to the mean) of the ionising continuum which quantify the amplitude of the flux fluctuations. We describe here the time-dependent effects identifying three different scenarios:
\begin{itemize}
\item{\it equilibrium state}: when $\teq \ll \tvar$ the plasma ionisation state is constantly in equilibrium with the ionising radiation. Under this condition, a large variation of the ionising continuum would lead to a significant change of the ionisation state. Either a very high density absorber or a slowly variable ionising source could lead to the equilibrium condition.
\item{\it non-equilibrium steady state}: when $\teq \gg \tvar$ the ionic concentrations do not respond to the ionization flux reaching a steady state defined by the ionisation and recombination states. The plasma is therefore constantly out of equilibrium with the ionising continuum. However, if the amplitude of the source variability is small, the differences from the mean are modest. This steady-state condition is common in scenarios with either very low-density absorbers, or a rapidly variable ionising source, or both. 
\item{\it delayed state}: when $\teq \sim \tvar $ the plasma ionic concentrations evolve smoothly and with a delay relative to the ionising luminosity variability. The plasma is neither in equilibrium nor in a non-equilibrium steady state. This is in principle the most interesting case since the time delays between the ionisation state of the plasma and the ionising luminosity is effectively used to constrain the density of the ionising source.
\end{itemize}
 
\noindent
Consequently, a time-dependent photoionisation analysis of a plasma in equilibrium-state or in non-equilibrium steady state will only allow us to derive, respectively, a lower and an upper bound on the gas density, which correspond to an upper and lower limit on its distance, respectively. Instead, observing the time delay on which the absorbing outflow responds to the ionising continuum will allow to constrain its location more precisely.

In order to understand which case applies to a plasma it is crucial to estimate both \teq and \tvar. In general, the power spectral density of fluctuations or the excess variance of the light curve \citep[e.g.,][]{Nandra97,Ponti12} help to determine the variability of the ionising continuum. The estimate of \teq is even more complicated since according to Equation \ref{eq:trec} it requires the knowledge of the density and ionisation state (thus the ionic concentrations, $n_{X,i}$) of the absorbing plasma. In Figure \ref{fig:trec}, we show the recombination timescale, which is a good approximation of \teq, as a function of the gas density assuming a typical type-1 AGN ionising continuum. In specific, we adopt the ionising spectral energy distribution (SED) of Mrk~509 taken from \cite{Mehdipour11} to compute the recombination timescales. In the {\it top} panels we display the \trec of all the relevant ions of oxygen and iron for a plasma with $\log \xi = 2.0$. The recombination timescale of different ions can vary by orders of magnitude from one ion to the next, even for ions of the same element. Different elements reach their equilibrium on different timescales.

In the {\it bottom} panels of Figure \ref{fig:trec} we show how the recombination timescale of a single ion (\oviii and \fexxvi) relates to the ionisation state of the plasma using a grid of $\log \xi$ spanning between -1 and 6. The ionisation level of the plasma leads the concentrations of the ionic species and consequently their recombination timescales. When the ionic concentrations reaches their maximum (in this case, $\log \xi \sim 2.0 $ and $\log \xi \sim 4.0$ for \oviii and \fexxvi, respectively) the ions require a longer time to reach the equilibrium. Moreover, the fact that we observe a larger scatter for \oviii with respect to \fexxvi is due to their different ionic concentration distribution: \oviii has a broader distribution than \fexxvi \citep[see for example Figure 6 of][]{Mehdipour16}.

Finally, when characterising a time-evolving ionised plasma it is important to include the cooling timescale, which indicates the time necessary for a gas to cool down. As we will show in the following section, the cooling timescale has a significant impact on the time evolution of the concentration of the plasma ions.

\section{A new time-dependent photoionisation model}
\label{sec:tdp}
Commonly-used photoionisation models are limited by the assumption that the plasma is constantly in ionisation and excitation equilibrium. Therefore, they are only suitable to describe the absorbers found in the $equilibrium$ case when $\teq \ll \tvar$. Using the standard photoionisation models to characterise the other two scenarios described in Section \ref{sec:time_evo} would lead to a wrong conclusion since the ionisation state of the plasma is in non-photoionisation equilibrium. In order to correctly characterise such absorbers, we developed a time-dependent photoionisation model (\tpho) which accounts for the time-dependency of the ionizing and recombining processes that take place in a photoionised plasma together with all the relevant heating and cooling processes. The ultimate goal of this model is to determine the location of the AGN outflows and consequently their energetic properties (kinetic power, momentum outflow rate, outflow mass rate), which are crucial to understand the AGN feedback mechanism. 

In the following section, we describe the \tpho model in detail explaining the algorithm and its inputs and outputs. Subsequently, to illustrate the time-dependent effects, we apply the model to two didactic cases characterised by a sudden increase and decrease of the ionising luminosity (Section \ref{sec:step_up} and \ref{sec:step_dw}, respectively). This simple cases allow us to study how the temperature, the heating/cooling rates and the ionic concentrations evolves when the gas is not any more assumed to be in photoionisation equilibrium. We also compare the time evolution of close ions (e.g. \oviii and \oix) and ions of different elements (i.e. oxygen and iron) for different densities.

A more realistic case will be examined in Section \ref{sec:flare}, where we calculate the time-dependent effects introduced by a flaring light curve. This specific example highlights the three scenarios explained in Section \ref{sec:time_evo} and the density range in which the model is sensitive for a given time variability. Finally, we compute the time evolution of the X-ray transmission of the plasma and compare it with the equilibrium solution in Section \ref{sec:transmission} to understand the importance of the time-dependent effects in a time-resolved X-rays spectroscopy analysis.

\subsection{TPHO model}
\label{sec:tpho}
The \tpho model enables a realistic characterisation of the evolution of the ionisation state of plasma exposed to a variable photoionising source. The model has been implemented in the X-ray fitting code \spex\footnote{\url{https://spex-xray.github.io/spex-help/index.html}} \citep{Kaastra96} and released to the community with the 3.07 version of the software \cite{Kaastra22}. 

The code evaluates the time evolution of all ionic concentrations for the elements with an atomic number between 1 (H) and 30 (Zn). To do so, it solves the set of $N$ coupled ordinary first-order differential equations shown in Equation \ref{eq:diffeq}, which is analytically solvable only for $N=2$. We solve all these equations simultaneously using the subroutine \texttt{solcon} of \spex which is based on the work of \cite{Kaastra93}. The method consists of the calculation on the fly of the transition matrix which contains all the ionisation and recombination rates for a grid of temperatures and ionisation parameters. Eigenvalues and eigenvectors of these matrix are also calculated. The ionisation balance is computed at small steps according to the time evolution of ionisation and temperature in the plasma. Finally, the eigenvector decomposition and the calculated coefficients are used to evaluate the ionic concentrations at each time step.

The code calculates the evolution of several heating and cooling processes: photoionisation, Compton scattering and ionisation, Auger electrons, free-free absorption, collisional de-excitation, and external heating for the heating rates whereas radiative recombination, collisional ionisation, inverse Compton scattering, bremsstrahlung, collisional excitation, and dielectronic recombination for the cooling rates. Cooling is assessed using the time-dependent ion abundances. The difference between the total cooling and heating processes is used to determine the temporal evolution of the plasma temperature.

The \tpho model needs several inputs for the determination of the time-evolution of the ionisation balance. For the initial conditions of the plasma, the model adopts photoionisation equilibrium in order to start the evolution of the plasma. The spectral shape of the ionising continuum is also necessary to determine the photoionisation rates. The \tpho model can, like \pion, take the spectral energy distribution directly from the continuum components set by the user in \spex. Alternatively, the SED can be provided via an input file. The variability of the ionising luminosity is also required to determine the time-evolution of the ionic concentrations. The code uses the light curve of the ionising luminosity to interpolate the ionisation parameter. The last fundamental parameter necessary to solve the time-dependent ionisation balance equations is the density of the plasma, \nh, which is the parameter of interest, and can be either fixed or fitted to the data in the \tpho model. 

The main outputs of the code are the ionic concentrations of all the elements, the heating/cooling rates and the plasma temperature as a function of time. We plot the outputs of the model in the following sections for three simple cases highlighting their dependence on the gas density. The \tpho code then uses the obtained ionic concentrations to calculate the corresponding absorption and emission spectrum of the plasma at each time step. Thus, a time-resolved spectrum taken at time $t$, can be fitted by applying the corresponding \tpho spectrum that is calculated at time $t$.

\subsection{Light curve case: step-up function}
\label{sec:step_up}

\begin{figure}
   \centering
   \includegraphics[width=\hsize]{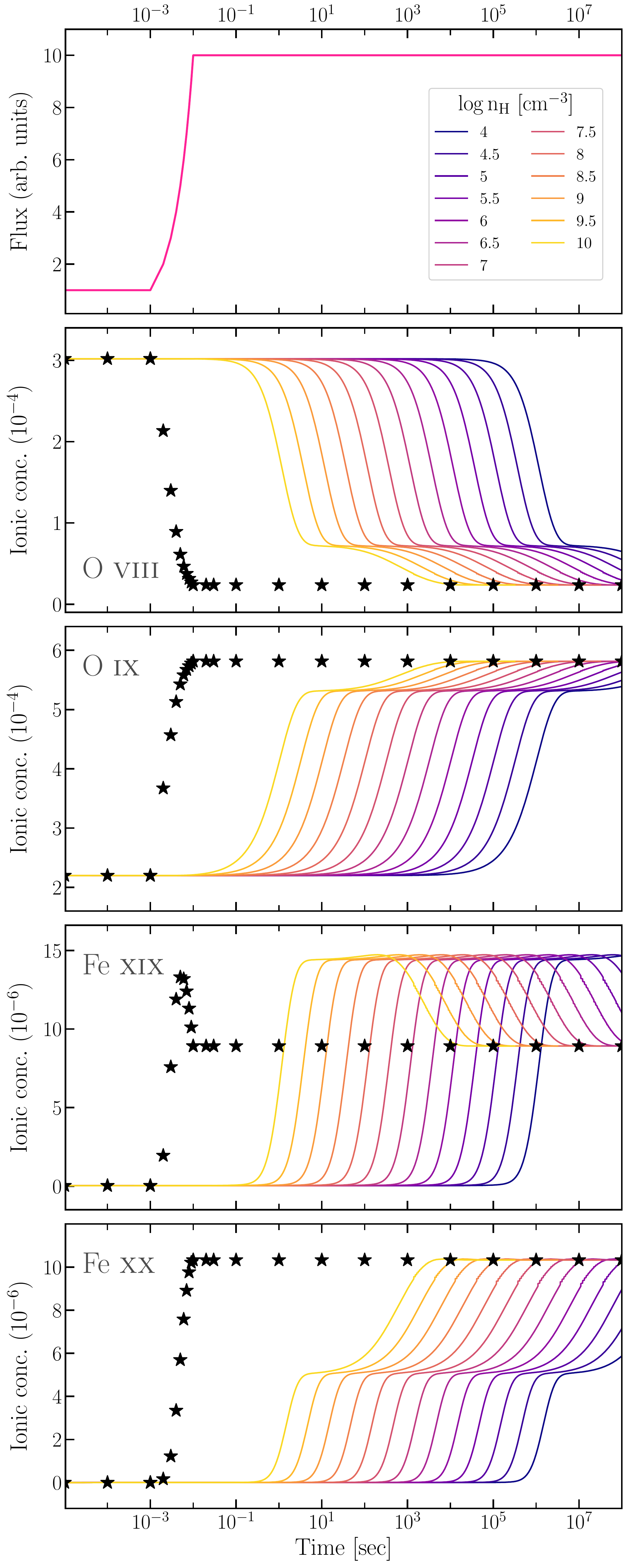}
      \caption{Step-up light curve case: the light curve is shown in the top panel. The ionising flux increases by a factor of ten in 0.01~s. In the panels below we show the time-dependent evolution of the concentration relative to hydrogen of \oviii, \oix, \fexix and \fexx for different gas densities and compared with the ionic concentrations for a plasma in photoionisation equilibrium (black stars).}
         \label{fig:stepup}
   \end{figure}

In order to examine the evolution of ionic concentrations calculated with \tpho, and compare them with the equilibrium scenario (\pion) we adopt the simplest case of a step-up function light curve. In this case, we assume that the ionising flux goes from a \textit{low} to a \textit{high} state almost instantaneously ($\Delta t \sim 0.01\ \rm s$) with a factor 10 change in flux. The light curve is shown in the upper panel of Figure \ref{fig:stepup}. For the initial condition we considered an optically thin gas cloud with an initial ionisation parameter of $\log \xi = 2$ and illuminated by the SED of Mrk~509 \citep{Mehdipour11}. With the \tpho model we computed the time evolution of the ionic concentrations for a grid of hydrogen densities ranging between $10^4$ and $10^{10}\ \rm cm^{-3}$. 

The time-dependent behaviour of the concentration of \oviii, \oix, \fexix, and \fexx are shown in Figure \ref{fig:stepup}. For different hydrogen densities the ionic concentrations follows the same exponential evolution but they systematically shift over different timescales. The shift is driven by the different density of the gas. Plotting the ionic concentration as a function of $n_{\rm H}\cdot time$ the curves will perfectly overlap (see for example top panel of Figure \ref{fig:up_dw}).

Ions in dense gases (yellow shades) reach their equilibrium concentration faster than the ions in less dense gases (blue shades). For comparison, we also calculated the ionic concentration curve for a gas in photoionisation equilibrium using the pion model (black stars). Interestingly, the evolution of the ionic concentration curves calculated with \tpho appears different from the equilibrium one. All the four ions show a flattening in their evolution which is closely related to the time-evolution of plasma temperature.

\begin{figure}
   \centering
   \includegraphics[width=\hsize]{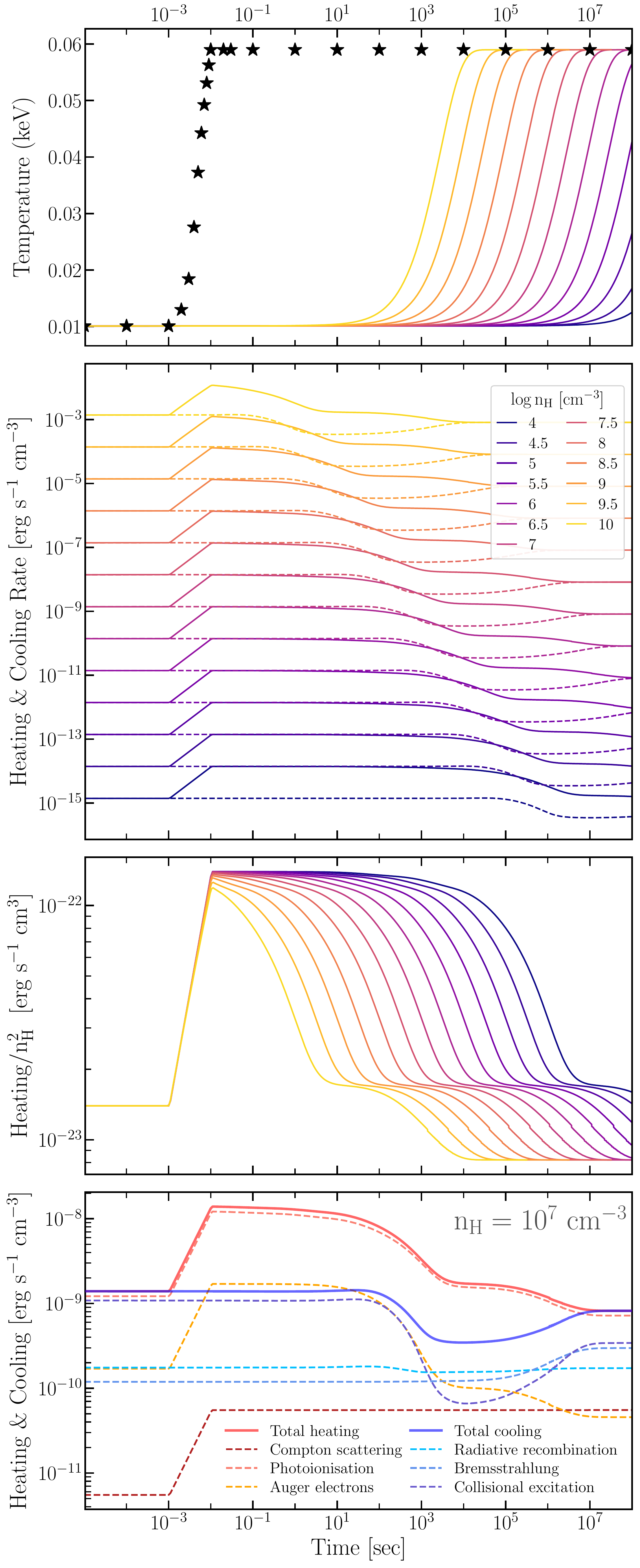}
      \caption{Step-up light curve case: the first two panels show respectively the time-dependent evolution of the electron temperature and the heating and cooling (solid and dashed lines) rates for a grid of density values. In the third panel, we compare the shape of the heating curves dividing them by its dependency on the gas density. The bottom panel illustrates the contribution of the main heating and cooling processes for a plasma with $\nh = 10^{7}\ \rm cm^{-3}$.}
         \label{fig:stepup_heat}
   \end{figure}

In the {\it top} panel of Figure \ref{fig:stepup_heat}, we illustrate how the temperature of the gas evolves considering the same step-up light curve described above. When considering the time-dependent effects the gas reaches the equilibrium temperature after a significant time delay with respect to a gas in ionisation equilibrium (black stars). This time delay increases with decreasing gas density. High-density gas with $\nh = 10^{10}\ \rm cm^{-3}$ takes $\sim10\ \textrm{ks}$ to reach the equilibrium temperature whereas low-density gases ($\nh < 10^6\ \rm cm^{-3}$) can take a few years ($t>10^{5}\ \rm ks$) to reach the equilibrium value. 

The plasma temperature is evaluated summing the total heating and cooling rates shown in the second panel of Figure \ref{fig:stepup_heat}: the temperature increases when the heating is larger than the cooling and is constant when the two processes become equivalent. The total amount of energy injected into the plasma strongly depends on the density of the plasma. The heating curves (solid lines) increase during the jump from low to high flux level. As soon as the ionising radiation reaches the high state, the heating rate in high density gas starts to drop. The cooling rates (dashed lines) start out as constant, but kick in at late times. These late changes in the heating and cooling rates, alongside temperature, naturally alters the ionic concentrations, making them rise or decline depending on the ion. Indeed, at around the time of the second decline/rise in ionic concentration shown in Figure \ref{fig:stepup}, there are noticeable deviations in the total cooling/heating rates.

In the third panel of Figure \ref{fig:stepup_heat}, we compare the shapes of the heating-rate curves which were computed fixing the initial ionisation parameter of the plasma. The curves show an extended peak close to their maximum heating rate value. The duration of this peak anti-correlates with the gas density. This shows again that denser plasmas react more quickly to incoming radiation. In the {\it bottom} panel, we show the contribution to the total of the different heating and cooling processes considered in the calculations for a gas with a density of $\nh = 10^{7}\ \rm cm^{-3}$. For the considered ionising continuum, the heating is dominated by photoionisation. Auger electrons and Comptonisation contribute less ($\sim 10\%$ and $<1\%$, respectively) and all the rest of the processes listed in the beginning of Section \ref{sec:tpho} can be neglected here. Collisional excitation, radiative recombination and bremsstrahlung are, instead, the major cooling processes. 

 
   \begin{figure}
   \centering
   \includegraphics[width=\hsize]{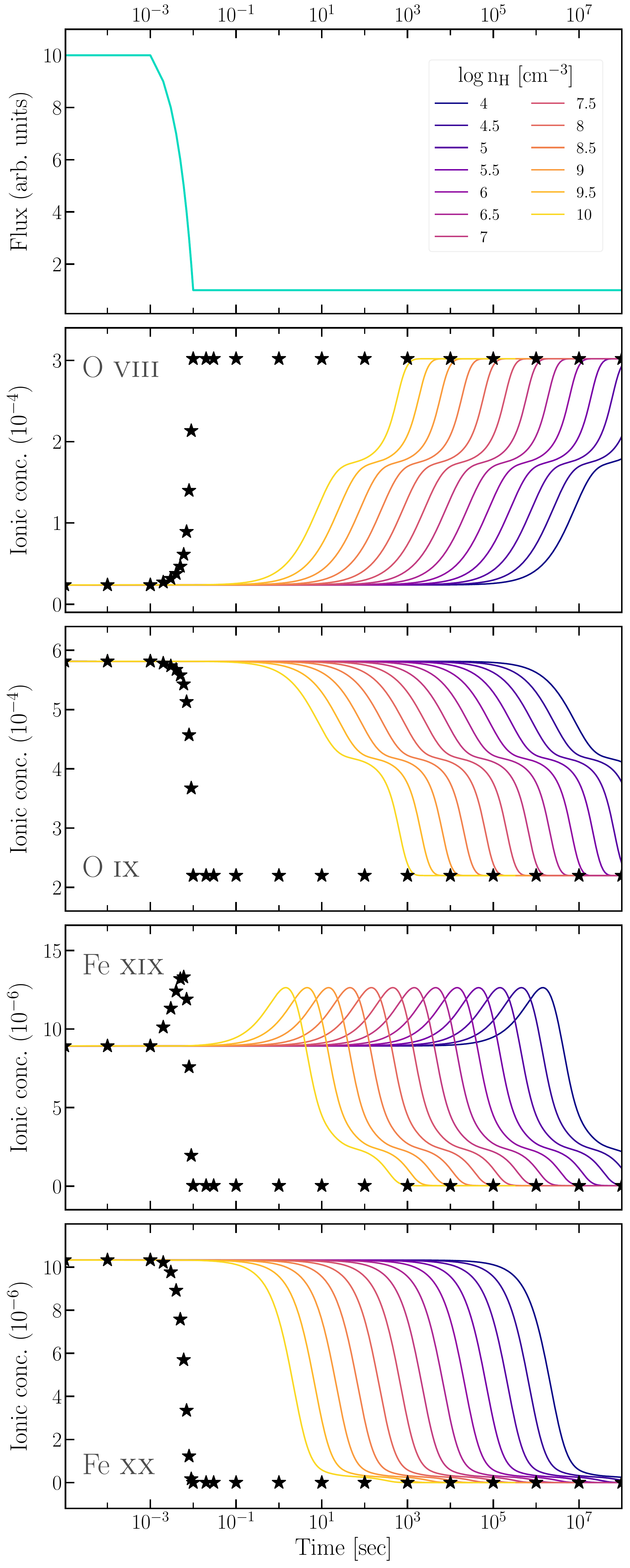}
      \caption{Step-down light curve case: the light curve is shown in the top panel. The ionising flux drops by a factor of ten in 0.01~s. In the panels below we show the time-dependent evolution of the concentration relative to hydrogen of \oviii, \oix, \fexix and \fexx for different gas densities and compared with the ionic concentrations for a plasma in photoionisation equilibrium (black stars).}
         \label{fig:stepdw}
   \end{figure}

 
   \begin{figure}
   \centering
   \includegraphics[width=\hsize]{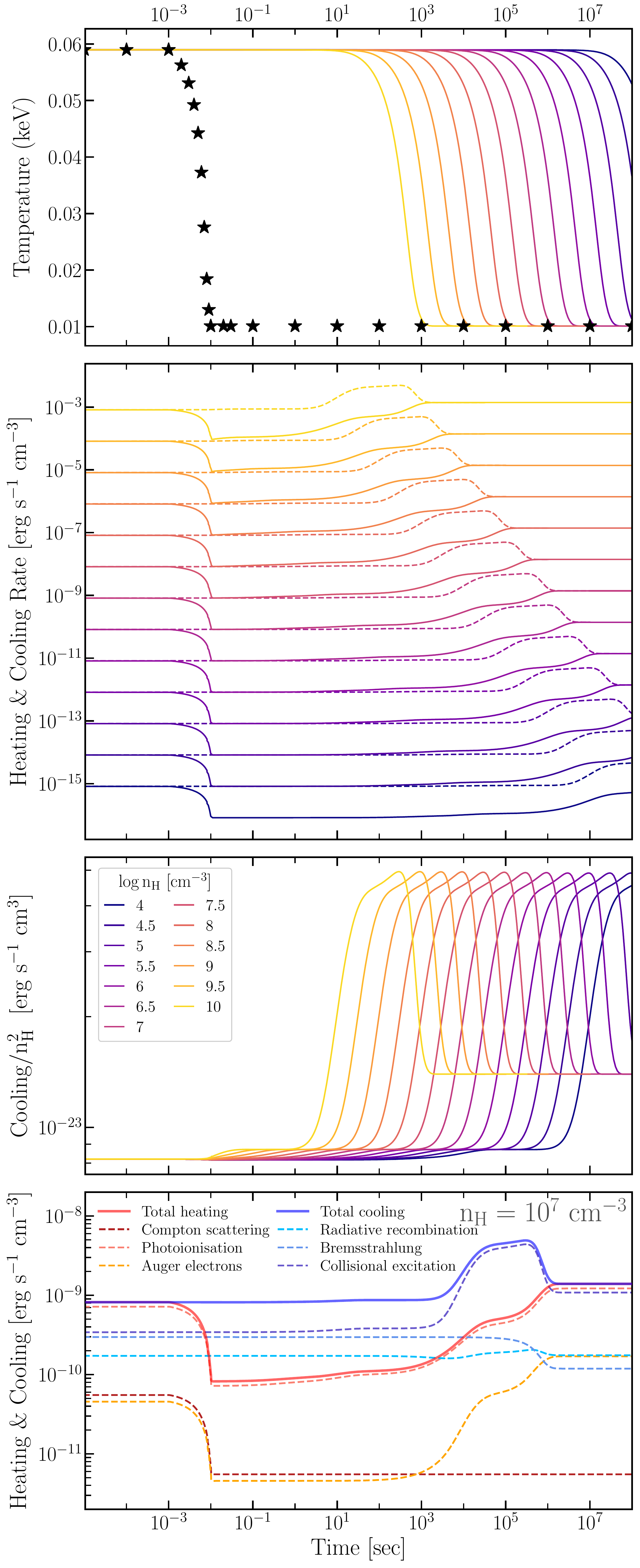}
      \caption{Step-down light curve case: the first two panels show respectively the time-dependent evolution of the electron temperature and the heating and cooling (solid and dashed lines) rates for a grid of density values. In the third panel, we compare the shape of the cooling curves dividing them by its dependency on the gas density. The bottom panel illustrates the contribution of the main heating and cooling processes for a plasma with $\nh = 10^{7}\ \rm cm^{-3}$.}
         \label{fig:stepdw_cool}
   \end{figure}

\subsection{Step-down light curve}
\label{sec:step_dw}

To compare the previous study case with a scenario where the cooling rates and the recombination processes are expected to dominate, we repeated the same investigation but for a step-down light curve function: the ionising continuum goes from high to low flux level by a factor of 10 (see light curve in the $top$ panel of Figure \ref{fig:stepdw}). To reproduce exactly the opposite of the previous test we assumed the SED shape of Mrk~509 and an ionisation parameter of $\log \xi = 3$ as initial condition. 

In Figure \ref{fig:stepdw}, we show the time evolution of the \oviii, \oix, \fexix, and \fexx concentrations. As shown before, the time necessary for each ionic concentration to reach their equilibrium values depends on their density. However, the ionic concentration curves do not follow the reverse pattern of the step-up light curve case shown before (Section \ref{sec:step_up}), as it would be expected for a plasma in ionisation equilibrium. 

In the {\it top } panel of Figure \ref{fig:stepdw_cool}, we show how the plasma temperature drops as a function of time and density. As shown before, the temperature is driven by the difference between cooling (dashed lines) and heating rates (solid lines) shown in second panel. In this case, the cooling is larger than heating and leads the temperature decrease. During the transition from high to low flux level, the cooling does not vary whereas the heating rate drops. As observed in the previous case, the change of the heating always happens at the same time independent of the gas density. After this jump, both cooling and heating rates stay steady for a time period that depends on the gas density: the lower the density the longer the steady phase. Then, both cooling and heating increase in the same manner until the cooling finally drops, reaching the same level of the heating. From this moment on, the gas is in photoionisation equilibrium with the ionising source.

The cooling rate curve has the same shape for different densities, but they are spread over different timescales, as shown in the third panel of Figure \ref{fig:stepdw_cool}. In the bottom panel, we show the total cooling rate curve for a gas with density of $\nh = 10^{7}\ \rm cm^{-3}$ with all the processes that contribute to it. Collisional excitation is the most important phenomenon and it is responsible for the bump observed at a later time in the total cooling curve. The ratios between the different contributions do not vary among the grid of densities considered here. Photoionisation, Compton scattering and Auger electrons are, instead, the main heating processes. 

\begin{figure}
   \centering
   \includegraphics[width=\hsize]{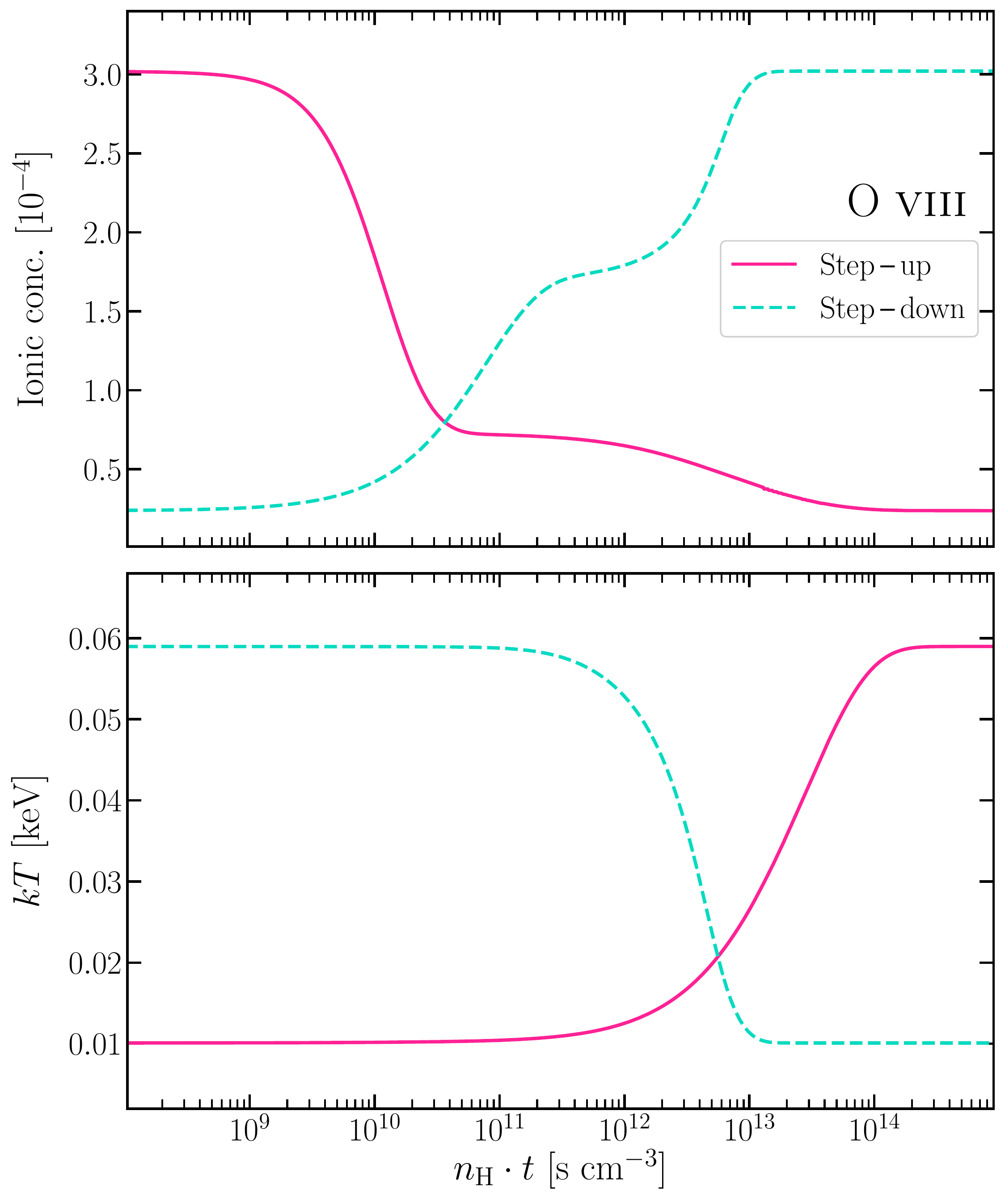}
      \caption{Comparison between the step-up and step-down scenarios. In the {\it top} panel, we show the evolution of the \oviii concentration as a function of $n_{H}\cdot t$. In the same rest-frame, the evolution of the plasma temperature is illustrated in the {\it bottom} panel.}
         \label{fig:up_dw}
   \end{figure}

It is now possible to compare the ionisation case, represented by the step-up function, with the recombination case, described by step-down function. In the {\it top} and {\it bottom} panel of Figure \ref{fig:up_dw}, we juxtapose, respectively, the evolution of the \oviii concentrations and the plasma temperatures as a function of $n_{\rm H}\cdot time$. The two concentration curves follow shapes which are not the inverse of each other, as we might expect to see for a gas in photoionisation equilibrium. They both include a plateau phase which, as we have already seen in Figure \ref{fig:stepup} and Figure \ref{fig:stepdw}, is common to all the ionic species. 

In the ionisation case (solid purple line), the concentration of \oviii reaches the plateau ten time faster than in the recombination configuration (dashed light-blue line). 
The duration of the plateau and the following rise (or drop) of the ionic concentration depends on the late evolution of the temperature. In the recombination case, the temperature varies significantly quicker than the ionisation case and it reaches the equilibrium value in a shorter time. Consequently, the recombining ionic concentration shows a shorter flat phase and reaches the final equilibrium state around ten time faster than the ionising curve. This differences show the importance of including the computation of the plasma temperature for an accurate time-dependent modelling of a photoionised plasma.

 
   \begin{figure}
   \centering
   \includegraphics[width=\hsize]{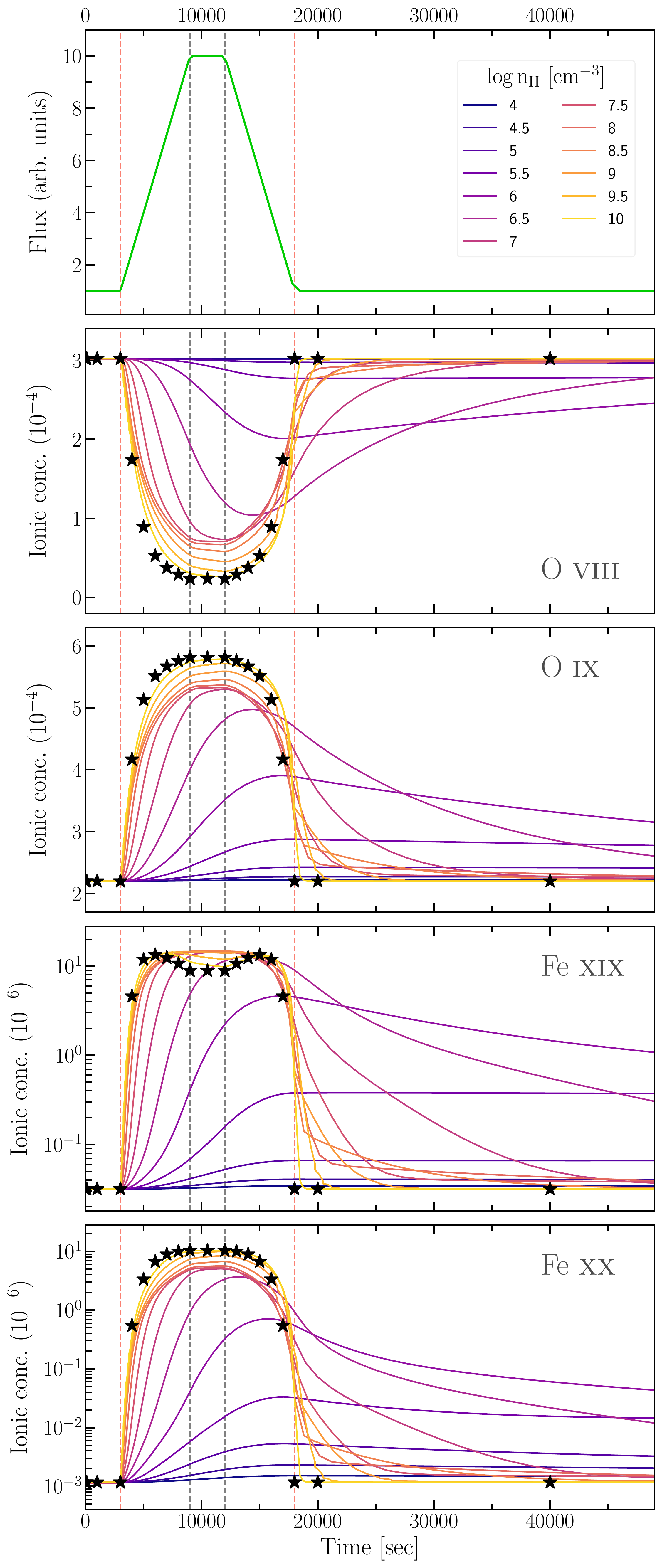}
      \caption{Flaring light curve case. The light curve is shown in the top panel. The ionising flux symmetrically increases and decreases by a factor of ten in 3~ks. The high state phase between the dashed grey vertical lines is 2~ks long. The onset and the end of flare are marked with two dashed red vertical lines. In the middle and bottom panels we show the time-dependent evolution of the concentration relative to hydrogen of \oviii, \oix, \fexix and \fexx for different gas densities and compared with the ionic concentrations for a plasma in photoionisation equilibrium (black stars).}
         \label{fig:flare}
   \end{figure}

 
   \begin{figure}
   \centering
   \includegraphics[width=\hsize]{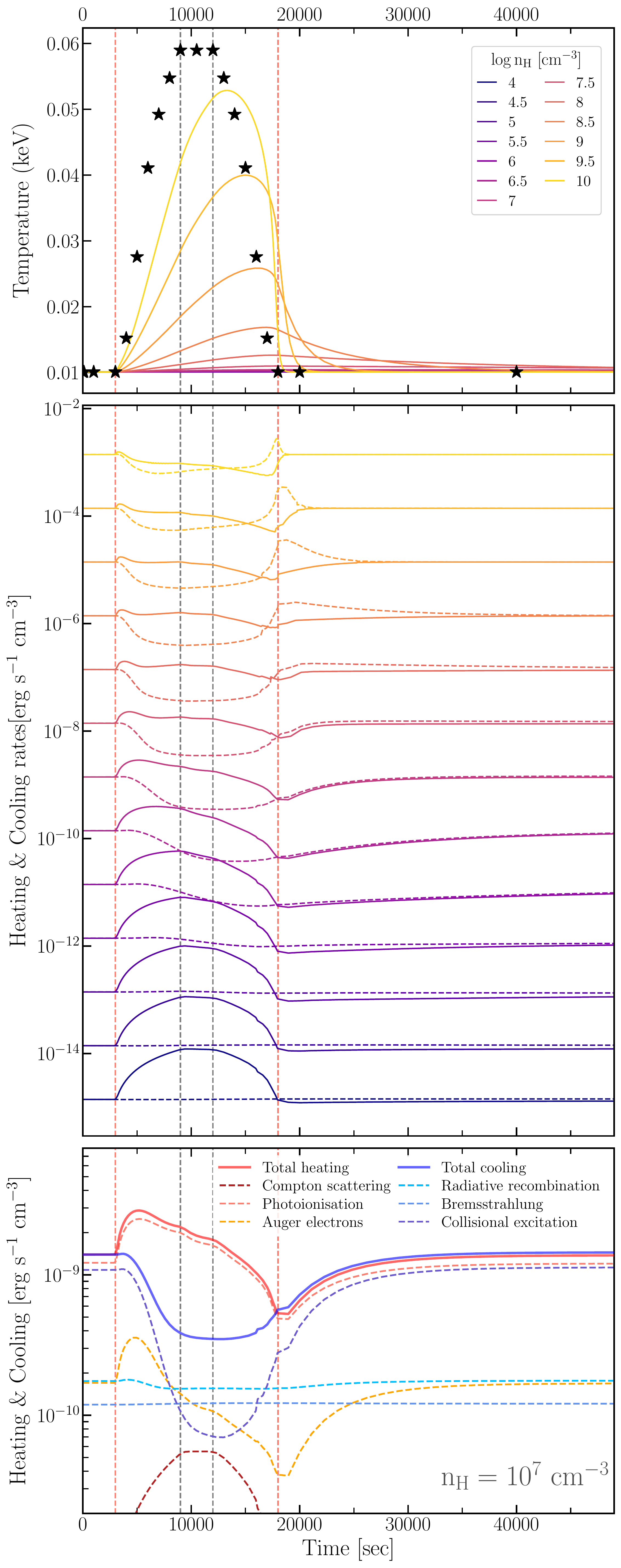}
      \caption{Flaring test case: the first two panels show respectively the time-dependent evolution of the electron temperature and the heating and cooling (solid and dashed lines) rates for a grid of density values. In the bottom panel we display the contribution of the main heating and cooling processes for a plasma with density $\nh = 10^{7}\ \rm cm^{-3}$.}
         \label{fig:flare_heatcool}
   \end{figure}

\subsection{Flaring light curve}
\label{sec:flare}

The previous two scenarios explained in Section \ref{sec:step_up} and \ref{sec:step_dw} represents two didactic frameworks that help demonstrate the time-dependent effects on the ionic concentrations. In this section we present a more realistic case where the ionising continuum flares briefly as shown in the {\it top} panel of Figure \ref{fig:flare}. In detail, the ionising luminosity increases by a factor of 10 in 6~ks (between the first red and grey vertical dashed line), stays in high state for 3~ks (between the two grey dashed vertical lines) and subsequently decreases to low state again in 6~ks (between the second red and grey dashed vertical lines). 

Using this flaring light curve, the SED shape of Mrk~509 and ionisation parameter $\log \xi = 2.0$, we computed the time evolution of the ionic concentrations. In Figure \ref{fig:flare}, we show the concentration curves of \oviii, \oix, \fexix, \fexx as a function of time and density. It is easy to distinguish in this case the three scenarios described in Section \ref{sec:time_evo}: $1)$ in gas with density above $10^8\ \rm cm^{-3}$ (yellow and orange shades) the ionic concentrations follow closely their $equilibrium$ values (black stars); $2)$ when the density is between $10^6\ \rm cm^{-3}$ and $10^8\ \rm cm^{-3}$ (red-purple shades) the ionic concentrations are $delayed$ and smooth with respect to the equilibrium curve; $3)$ the ionic concentrations of low-density gases, with $\nh < 10^{6}\ \rm cm^{-3}$ (blue shades) are $steady$ and do not vary significantly with time. The gas does not have enough time to respond to the initial increase of the ionising luminosity. 

The time evolution of \fexix concentration follows a double-horn curve which differs from the bell shape of the other three ions illustrated. Its ionic distribution peaks during the increase and decrease phases, between which the ionic state becomes less populated in favour of \fexx, as the ionizing flux reaches its maximum. We already saw this transition in the third plot of Figure \ref{fig:stepup} and \ref{fig:stepdw} for the increasing and decreasing luminosity phase, respectively.

The time evolution of the temperature is shown in the {\it top} panel of Figure \ref{fig:flare_heatcool}. Only high-density plasmas (with $\nh > 10^{8}\ \rm cm^{-3}$) significantly increase their temperature during the flare of the ionising luminosity. Moreover, the temperature peaks at a delayed time with respect to the flare. In the {\it middle} panel, we compare the total heating and cooling for the considered density grid. In this flaring scenario, the time-evolution of the heating and cooling strongly depends on the gas density. For instance, the cooling is steady for low-density gas, whereas it quickly varies in the densest plasmas. Instead, the heating rate starts to evolve similarly to the ionising radiation as soon as we consider low density gas. Even though we mimic a light curve where the ionising flare is symmetric, we see that during the flare the heating time is longer than its cooling time and dominates in gas with medium and low density ($\nh < 10^{8}\ \rm cm^{-3}$). To reach their energetic equilibrium (i.e. $\rm colling=heating)$, the low density gases require a longer time than the time interval illustrated here. This explain why the cooling and heating do not match at the right end of the plot. In the $bottom$ panel, we show the main contribution of the different processes to the heating and cooling rates. Similarly to the step-up and step-down cases studied before, photoionisation is the main heating process, whereas collisional excitation together with radiative recombination and bremsstrahlung are the most relevant cooling phenomena.

 
   \begin{figure*}
   \centering
   \includegraphics[width=\hsize]{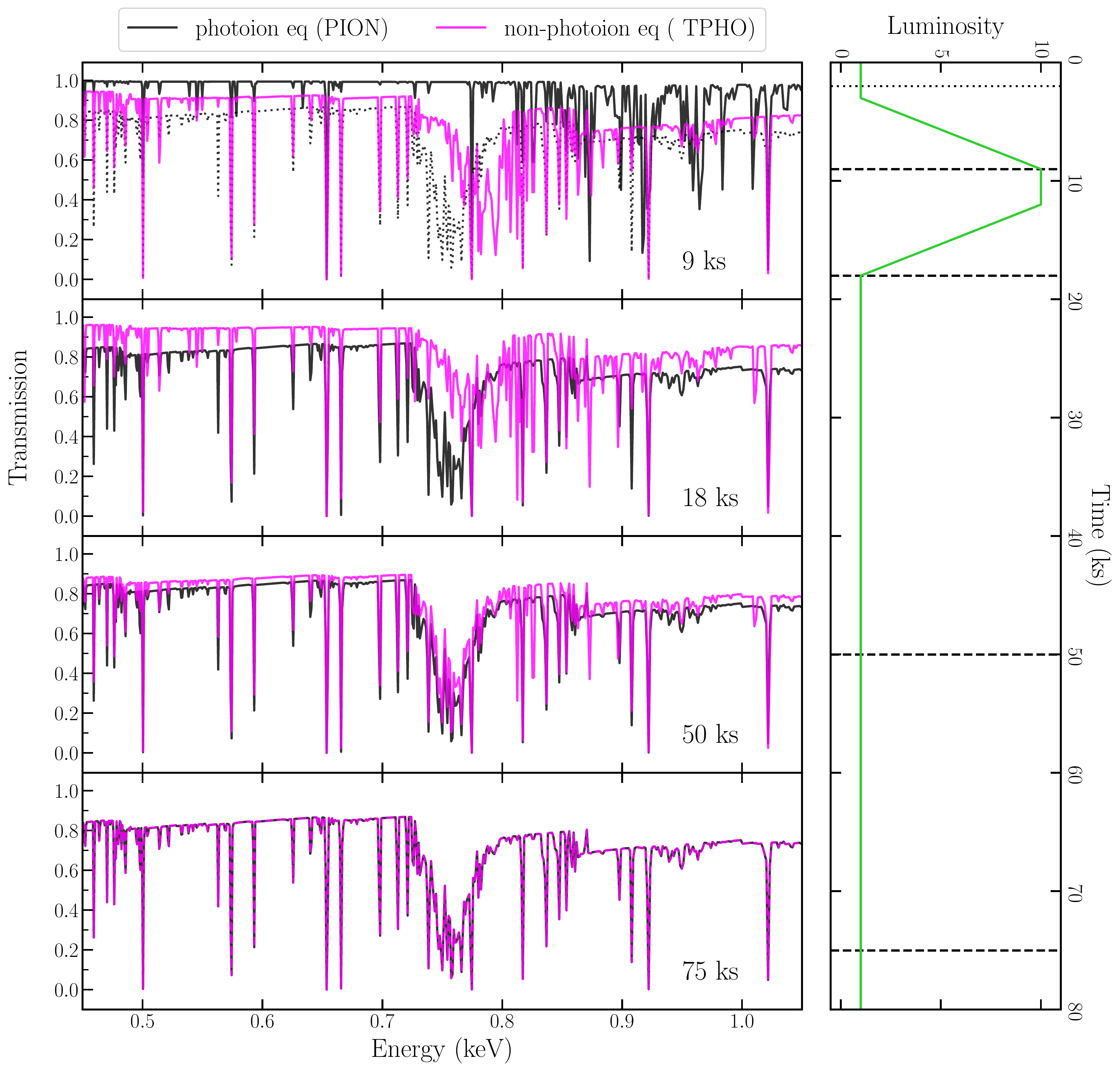}
   \caption{Comparison between the X-ray transmission (line and continuum absorption) of a plasma in photoionisation equilibrium (black solid line) and in non-photoionisation equilibrium (magenta solid line). We assume a plasma with a density of $n_{\rm H}=10^6\ \rm cm^{-3}$ and an initial ionisation parameter of $\log \xi = 1.8$. We show the comparison at four different epochs: from top to bottom 9~ks, 18~ks, 50~ks and 75~ks which correspond to the black dashed lines in the light curve panel on the right. In the top panel we also plot the initial transmitted spectrum (black dotted line), right before the onset of the flare.}
   \label{fig:transmission}
   \end{figure*}

\subsection{Transmitted spectrum}
\label{sec:transmission}

 
   \begin{figure}
   \centering
   \includegraphics[width=\hsize]{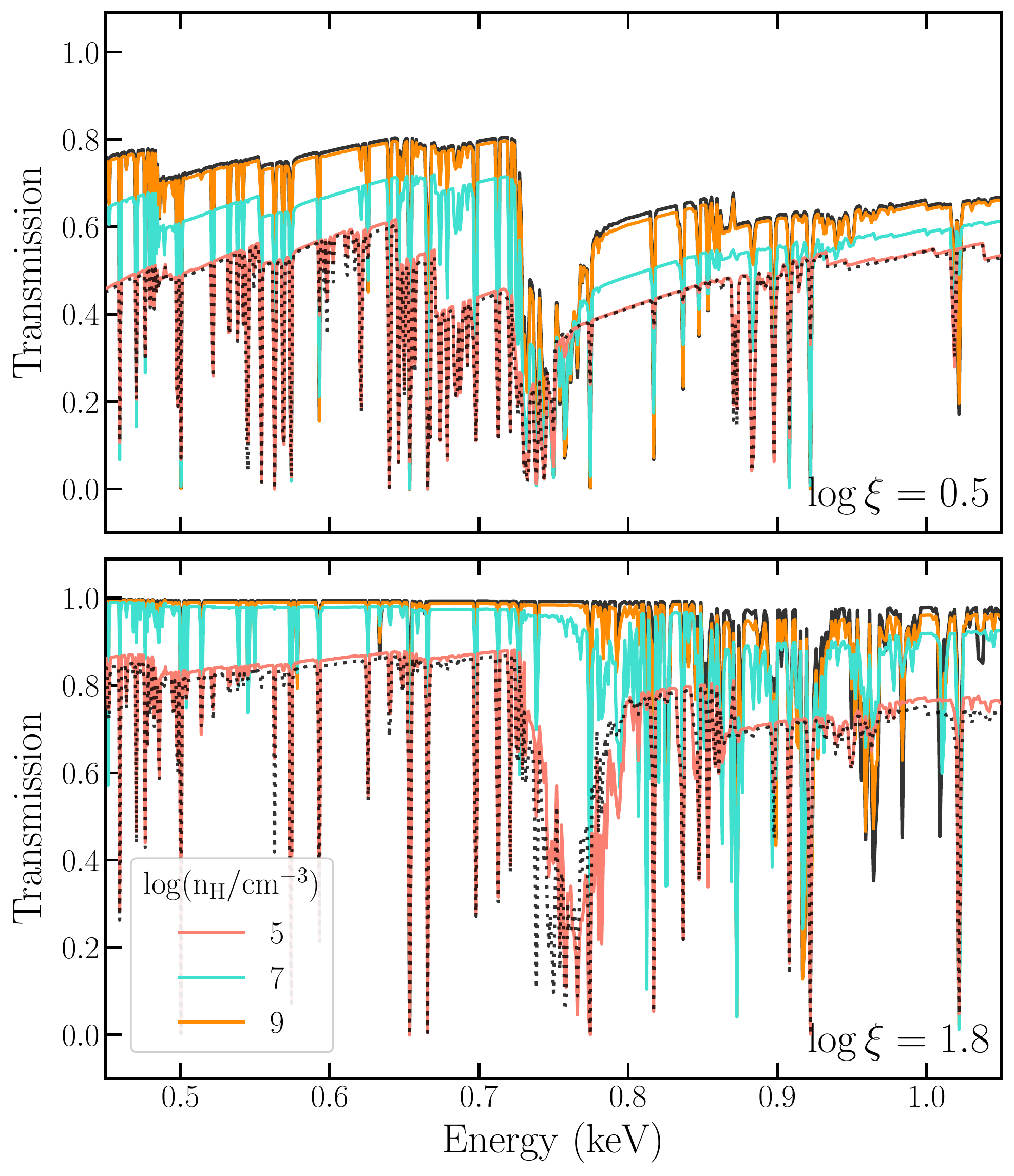}
   \caption{X-ray transmission of an absorber assuming different densities ($\rm 10^{5} \ \rm cm^{-3}$ in red, $\rm 10^7 \ cm^{-3}$ in turquoise, and $\rm 10^9 \ \rm cm^{-3}$ in orange) and initial ionisation parameters ($\log \xi = 0.5$ top panel and $\log \xi = 1.8$ bottom panel). We extracted all the spectra at the peak of the flaring light curve ($t=9 \rm \ ks$, see Figure \ref{fig:transmission}). In both panels we overplot the initial transmitted spectrum ($t=0$, black dotted line) and the spectrum of a plasma in photoionisation equilibrium with the ionising source at $t=9~\rm ks$ (black solid line).}
   \label{fig:trans_den}
   \end{figure}

In the final step of the \tpho computation, all the ionic concentrations are used to build the transmitted spectrum of the plasma. It comes as a multiplicative component and it can be used to characterise the evolution of the absorption features observed in the data. The code provides a transmitted spectrum at each desired point of the light curve, e.g. when the X-ray spectrum of the source is available. Moreover, \spex allows a simultaneous fit of multiple epochs, which helps to obtain stronger constraints on the density of the absorber. 

The time-dependent effects influence the spectral shape of the X-ray transmission as a function of time. In Figure \ref{fig:transmission}, we compare the evolution of the transmitted spectrum for a plasma in photoionisation equilibrium (in black), calculated with the \pion model and the one calculated with the new model \tpho, which takes into account the time dependent effects (in magenta). We illustrate the case of the flaring light-curve (see Section \ref{sec:flare}).
For the calculation, we assumed an initial photoionisation equilibrium with an ionisation parameter of $\log \xi = 1.8$ (the initial transmitted spectrum is shown with the black dotted line in the top panel), a column density of $\NH = 5\times10^{21}\rm \ cm^{-2}$, and a density of $\nh = 10^6 \rm \ cm^{-3}$ and the SED of Mrk~509. This combination of assumption gives a system in the {\it delayed} state option. We extrapolated the spectra at four different intervals of the light curve shown on the right panel: in particular at the flare maximum (top panel), right after the flare (middle panel) and after a long period of steady-state (two bottom panels).

The comparison highlights the impact of the time-dependent effect on the X-ray transmission spectrum. The deviations from the equilibrium spectrum are larger during the flare activity when the flux varies rapidly. The plasma slowly responds to the rapid increase of the ionising luminosity and it shows a lower ionisation state at the peak of the flare. The shape of the iron unresolved transitions array which produces prominent features in the $730-830$~eV energy range ($15-17$~\AA) and the overall opacity represent the strongest differences between the two transmitted spectra in the first two extraction epochs. Before the third extraction epoch, due to the steady flux, the gas has time to recover the photoionisation equilibrium and the two X-ray transmissions become similar. The gas requires a longer time ($t\sim75\rm\;ks$) to fully recover the equilibrium (last epoch).

The density of the plasma controls the evolution of the ionic concentrations (see Figure \ref{fig:stepup}, \ref{fig:stepdw}, and \ref{fig:flare}) and therefore the evolution of the X-ray transmission. The transmitted spectrum of lower density gases changes slower than the one of higher density plasma. In Figure \ref{fig:trans_den}, we compare the transmitted spectrum at the peak of the flaring light curve (9~ks) for several densities and two different initial ionisation states ($\log \xi = 0.5$ on the top panel and $\log \xi = 1.8$ on the bottom panel). Regardless of the increase of the intrinsic luminosity, the X-ray transmission of gases with $n_{\rm H} \lesssim 10^{5}\ \rm cm^{-3}$ does not diverge from the initial state (dotted black line). The equilibrium timescale for these absorbers is significantly longer than the increasing period and therefore the absorber does not have enough time to respond to the luminosity variation. In contrast, plasmas with $n_{\rm H} \gtrsim 10^{9}\ \rm cm^{-3}$ respond almost simultaneously to the increase of the luminosity and they can be considered in photoionisation equilibrium (black solid line). Finally, the opacity of the plasma decreases significantly with increasing density. High-density gases can, indeed, reach higher ionisation states due to their shorter equilibrium timescales. 

 
   \begin{figure}
   \centering
   \includegraphics[width=\hsize]{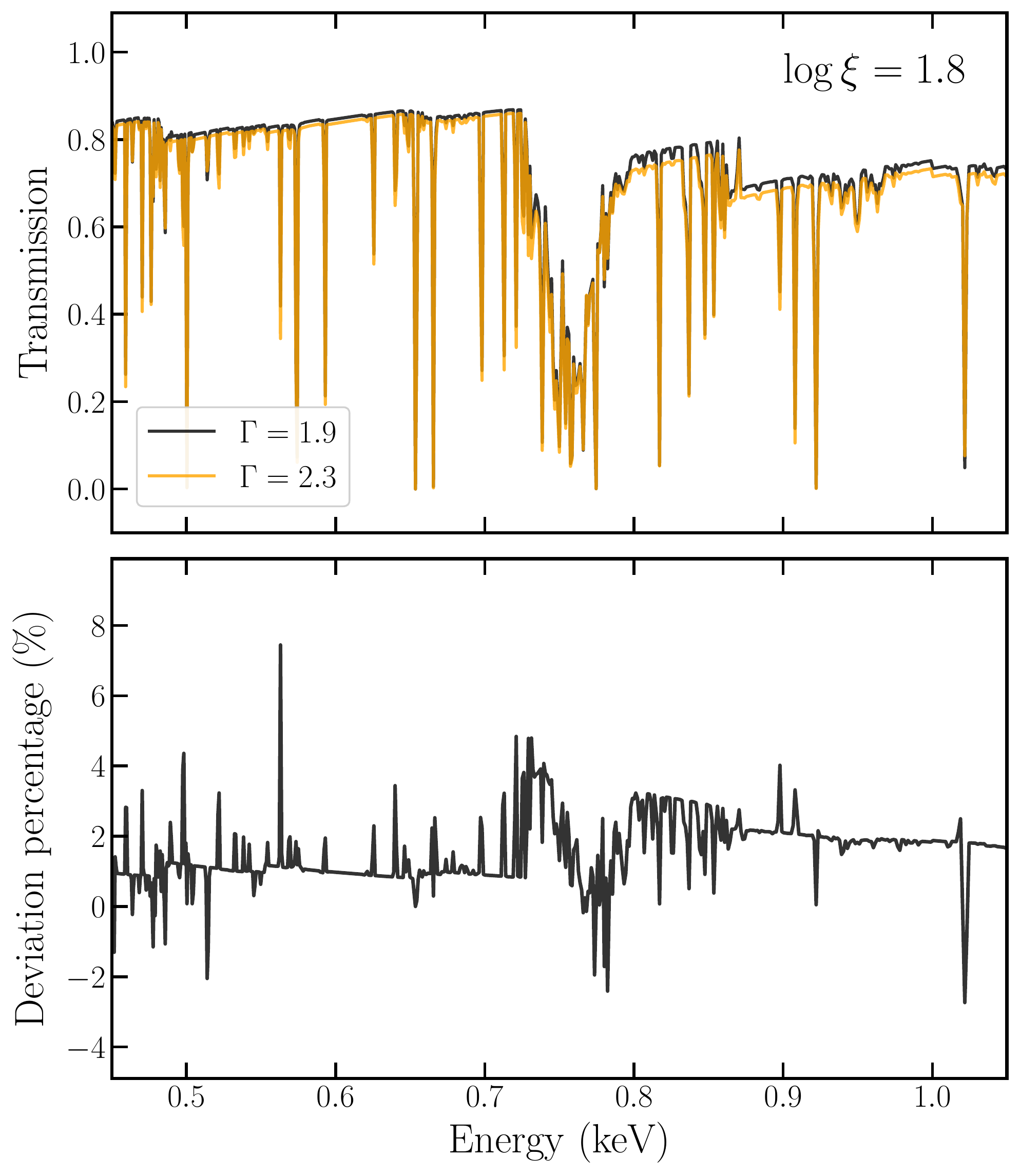}
   \caption{Impact of the ionising SED on the transmitted spectrum. In the top panel we compare the X-ray transmission of two plasmas with a different ionising SED and with the same initial ionising parameter and density: black line represents the transmittance for a plasma which does not vary the SED shape instead the yellow line shows the transmittance of a gas that sees an increase of the power-law index by $20\%$. The deviation between the two models is shown in the bottom panel.}
   \label{fig:sed_tr}
   \end{figure}

\subsection{Caveats}
\label{sec:limits}
At present, our time-dependent photoionisation model relies on a few assumptions. Firstly, the model considers the plasma in photoionisation equilibrium with the ionising luminosity at the starting point of the light curve ($t=0$). This assumption is necessary to evaluate the temperature, the initial ionic concentrations and heating and cooling rates of the plasma. However, it is possible that an AGN outflow, for example, it is already in non-equilibrium with the X-ray emission at the beginning of the observing campaign. This can introduce uncertainties on the response time of the plasma and consequently also on its inferred density. In order to limit this uncertainty, it is important to monitor the intrinsic luminosity for an extended time interval.The initial equilibrium can be placed after a long phase of steady flux where the ionised plasma is most likely in photoionisation equilibrium.

Moreover, we neglect any change of the SED components such as the power-law slope, extent of the soft excess, reflection, or the optical/UV bump during the considered time. The shape of the SED is kept frozen during the calculation of the ionic concentration evolution. Only the total normalisation varies in correspondence with the ionising luminosity. Therefore, the model is particularly suitable to describe the time-dependent effects of AGN that do not show a strong variability of the SED shape. 

Differences in the ionising SED have direct impact on the ionising balance and the thermal stability of photoionised plasmas \citep{Mehdipour16}. However, the changes in the transmitted spectrum are limited to small changes of the ionising SED. In Figure \ref{fig:sed_tr}, we compare the transmitted spectra considering (orange) and ignoring (black) any variation of the spectral shape. In detail, we let the power-law index, $\Gamma$ increasing by $~20\%$ (from 1.9 to 2.3) and calculated the transmitted spectrum at the flaring peak ($t = 9\ \rm ks$, see Figure {\ref{fig:transmission}}). We assumed a density of $\nh=10^6\ \rm cm^{-3}$ and an initial ionisation parameter of $\log \xi = 1.8$. The deviation between the two models is shown in percentage in the bottom panel. The discrepancies are within $~5\%$, which can be neglected with the current high-resolution X-ray spectrometers. In a future version of the code we will implement a more sophisticated algorithm able to take into account any change of the ionising SED.

We consider only optically thin slab of plasma where the propagation time of the X-ray radiation is smaller than both the source variability timescale and the associated recombination timescale \citep[see][]{Schwarz72,Binette88}. We neglect any time-dependence of the radiative transfer which becomes important for optically thick plasma \citep[see][]{Garcia13b}. Recently, \citet{Sadaula22} extensively studied the evolution of the ion fraction and temperature at various depths in the cloud solving the time-dependent radiative transfer equation. They show the relevance of this effect for clouds with a thickness of $10^{16-17}\ \rm cm$. We refer to their work for a detailed comparison between the time-dependent effects of ionisation balance and the ones of the radiative transfer. 

Finally, the present model assumes that all ions are in their ground state. This is good enough as far as total time-dependent ionization/recombination rates are computed. At high density many of the levels of ground terms and configurations become populated by collisions leading to population of metastable levels \citep{Kallman21}. The critical density leading to population of the excited levels is $\sim 10^{14} \ \rm cm^{-3}$ \citep{Mauche04}. For higher densities this would give obviously some deviations in the computed time-dependent ion concentrations, but not by large amounts since the main effect of the density is setting the proper time scales. Level populations are not computed, because that requires much more computational effort. Thus, for higher densities, the ion concentrations are reasonable, but obviously care should be taken with the absorption spectra (which are computed assuming all ions are in the ground state).

\section{Discussion}
\label{sec:discussion}

 
   \begin{figure*}
   \centering
   \includegraphics[width=\hsize]{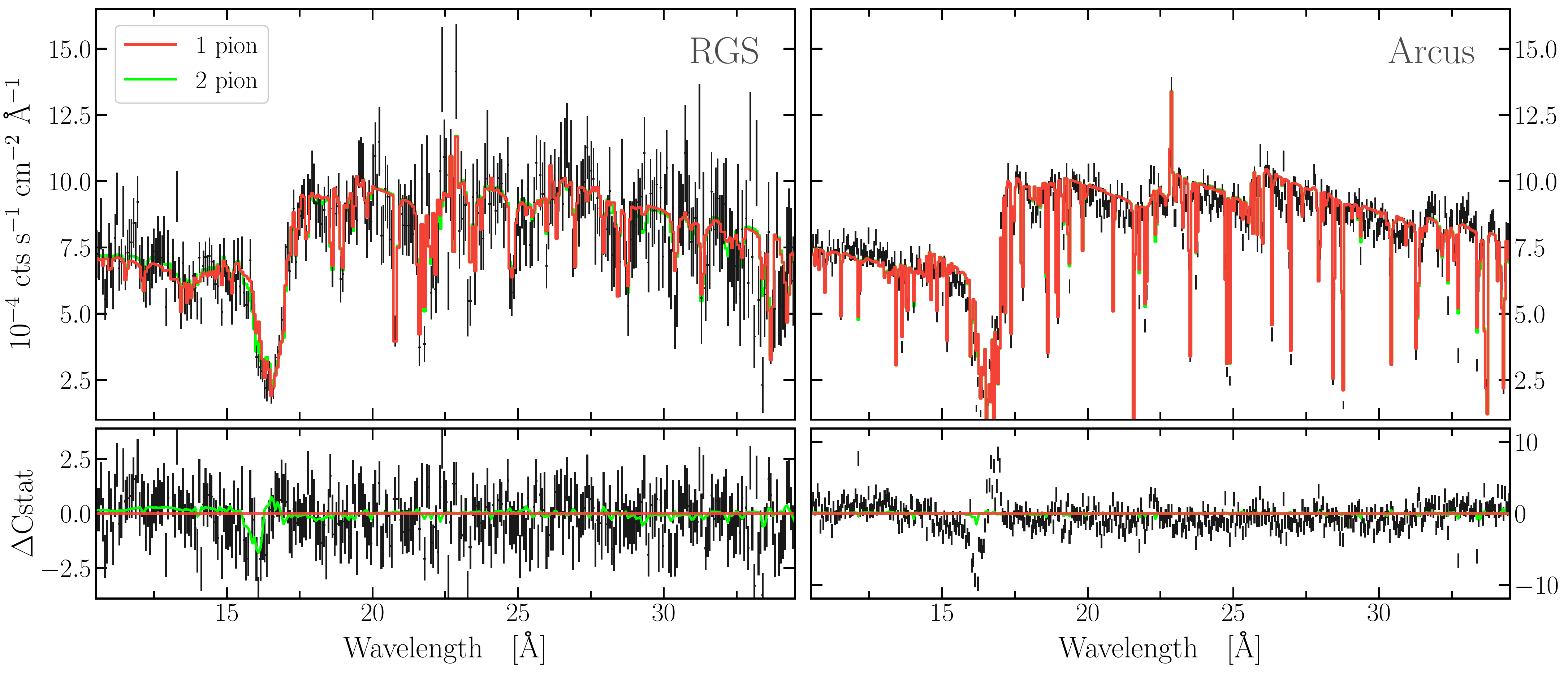}
   \caption{Time-dependent effects in a synthetic RGS and \arcus spectra of an AGN (SED from Mrk~509) with a warm absorber in absorption. To simulate the data we computed the time-dependent photoionisation model 25~ks after a sharp increase of the ionising luminosity. In red, we show the best fit using a \pion component and neglecting the time-dependent effects. The residuals are displayed in the bottom panels. The best fit using two \pion components is shown in green.}
   \label{fig:simu}
   \end{figure*}

 
   \begin{figure}
   \centering
   \includegraphics[width=\hsize]{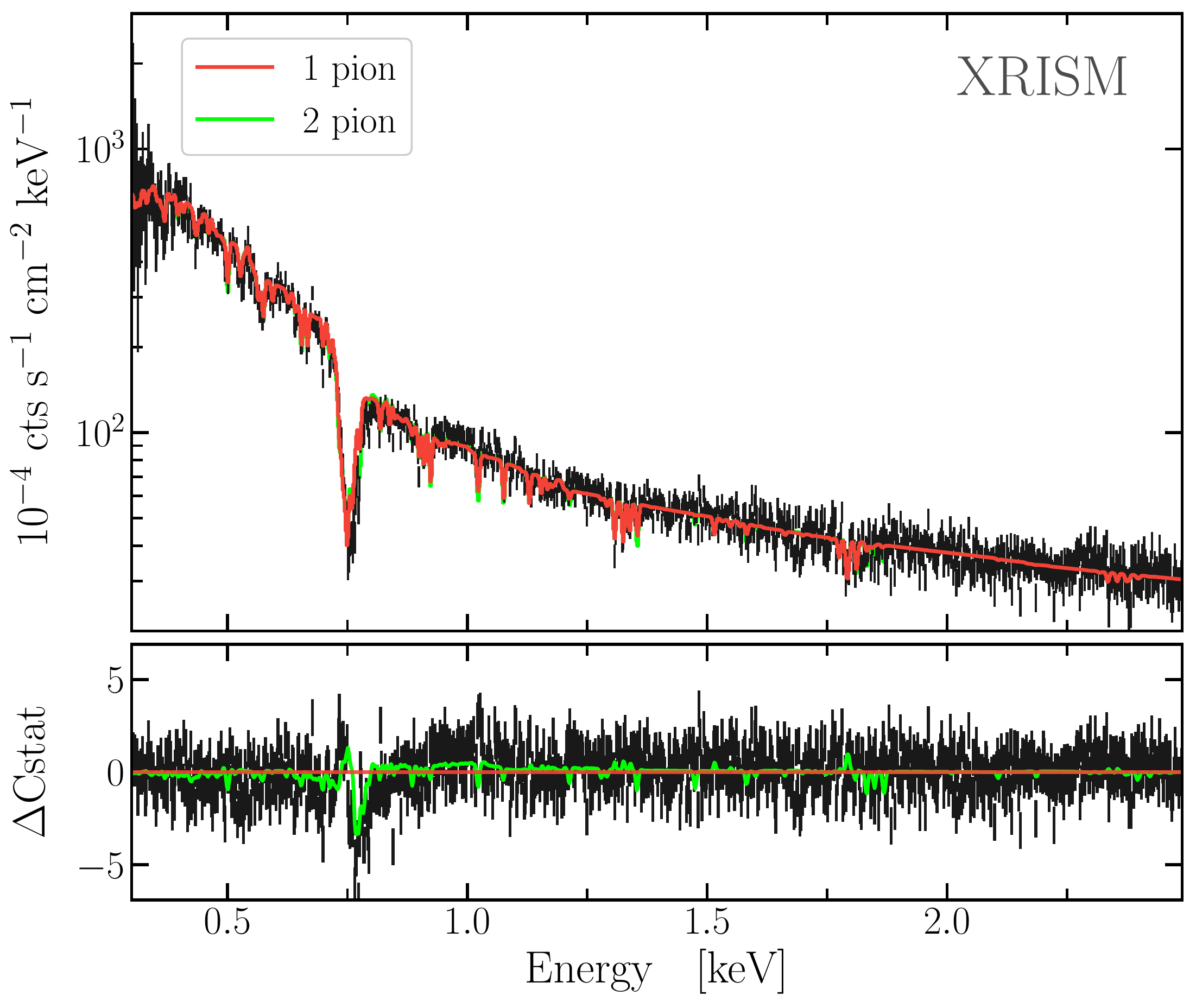}
   \caption{Time-dependent effects in a synthetic XRISM spectra of an AGN (SED from Mrk~509) with a warm absorber in absorption. To simulate the data we computed the time-dependent photoionisation model 25~ks after a sharp increase of the ionising luminosity. In red, we show the best fit using a \pion component and neglecting the time-dependent effects. The residuals are displayed in the bottom panels. The best-fit using two \pion components is shown in green.}
   \label{fig:xrism}
   \end{figure}

Standard photoionisation models such as \spex, \textsc{Cloudy} \citep{Ferland17} and \textsc{Xstar} \citep{Bautista01} allow us to characterise astrophysical plasma in ionisation equilibrium. This equilibrium assumption is valid in presence of a steady ionising luminosity or when modelling high-density plasmas. A sudden variation of the ionising luminosity can cause departure from the photoionisation equilibrium. The equilibrium timescale (see Equation \ref{eq:trec}) indicates the time necessary for the plasma to recover the ionisation balance. When this timescale is similar to or longer than the ionising source variability timescales the gas is non-equilibrium photoionisation with the ionising radiation. In other words, the assumption that heating and cooling rates are equal is not valid any more and time-dependent effects should be taken into account \citep[e.g.,][]{Krolik95,Nicastro99}. 

\tpho, our new photoionisation model based on \pion, includes the time-dependence of all ionising/heating and recombining/cooling processes in the plasma. The model is able to characterise the ionisation state of a plasma photoionised by a variable source. The full atomic database of \spex is used to calculate the evolution of all the ionic species by solving the differential Equation \ref{eq:diffeq}.

Earlier time-dependent photoionisation codes used only limited atomic data to study the evolution of absorption features of low-resolution X-ray spectra. For example \cite{Nicastro99} characterised the high signal-to-noise \rosat observation of the Seyfert I galaxy NGC~4051 omitting photoionisation from the L shell and all the iron ions. This introduced uncertainties on the computation of the abundance balance of low ionised gases and a possible overestimation of the O and Ne K shell photoionisation. \cite{Morales00} studied the complex time behaviour of the oxygen absorption features detected in the \asca and \bepposax spectra of MCG-6-30-15 using code developed by \cite{Reynolds96} limited to only oxygen ions. High-resolution \chandra/HETG \citep{Canizares05} and \xmm/RGS \citep{denHerder01} spectra require an extended atomic model in order to analyse the variability of the whole forest of absorption lines. To study the variability of the absorbers along the line of sight of Mrk~509 \cite{Kaastra12} used \textsc{Cloudy} to generate several time-dependent photoionisation models for a grid of densities. 

By implementing our \tpho model for the \spex package we do not only have direct access to the \spex database and photoionisation routines, but we can generate the transmitted spectrum which can be promptly multiplied with the broadband model. As a results, it is possible to directly fit the data and constrain the density of the absorber. The model is therefore suitable to study the time behaviour of the absorption features detected in the high-resolution spectra taken with current and future X-ray missions.

It is also possible to extend the time-dependent photoionisation modelling to high-resolution spectra in the UV band \citep[e.g.,][]{Arav20}. In case the absorber has spectral features in both the UV and X-ray band and both spectra are available, a simultaneous UV/X-ray analysis would accurately determine the time-dependent properties of the plasma. Finally, the \tpho model can be used to verify the density estimates from metastable lines in both UV and X-ray energy bands.

The primary scientific targets of the \tpho model are the outflows observed in bright Seyfert I galaxies. The code allows to study the time behaviour of both low and high ionisation plasma, tuning their flow and dispersion velocities, relative abundances, and column density. The main information we can access with the \tpho model is the density of the absorber. The time-dependent effects strongly depend on this quantity (see Section \ref{sec:time_evo}). Consequently, by reliably deriving the density of the plasma, it is possible to accurately locate the outflows and derive their energetics.

Moreover, the time variable spectra provided by \tpho can be used to predict the effect of warm absorber or obscuration events on the time lags and coherence of the energy dependent light curves. The response of the gas to changes in the ionising continuum can indeed introduce additional lags \citep{Silva16}. Thus, recognising the contribution of the recombining gas to the AGN X-ray time lags is crucial to interpret the continuum lags connected to propagation and reflection effects in the inner emitting regions. Recently, \cite{Juranova22} demonstrated that the gas response to the source radiation results in a decrease of the coherence in the Fourier timing analysis which can be used to constrain the gas density, opening therefore a new methodology to derive the location of outflows.

Tidal-disruption events (TDEs) and gamma-ray bursts (GRBs) represent two potential astrophysical systems where a time-dependent photoionisation modelling can be applied. In these transient events, the drastic increase of luminosity photoionises the surrounding gas (either ejecta or host galaxy); the explosive phase is then followed by a rapid decrease of the flux where the gas recombines on timescales determined by the density. The variability study of the absorption features in the spectra of such extreme events is, however, limited by the collecting area and the slew capability of current X-ray telescopes. Absorption lines are hardly detected even in the X-ray spectrum of the brightest GRBs \citep{Campana16} and only a few luminous TDEs show spectral features imprinted by an outflow \citep{Miller15}. Future X-ray missions, in particular {\it Athena}, will likely enable time-dependent photoionisation modelling of these extreme astrophysical events \citep{Piro21}. 

\subsection{Simulation}
\label{sec:simulation}

We discuss here the impact of the time-dependent effects on a standard time-resolved X-ray spectroscopy campaign of AGN outflows. In Figure \ref{fig:trans_den}, we demonstrate how the shape of the transmission spectrum of a warm absorber calculated with our \tpho model strongly diverges from the equilibrium one depending on the density of the plasma. We now investigate how well a standard photoionisation model can describe the spectral features of a gas in non-equilibrium with the ionising source. We aim to answer the question: is a time-dependent photoionisation model really necessary to fit the AGN outflow spectra?

To address this question we simulated the absorbed spectrum of an AGN after a sudden increase in the intrinsic luminosity by a factor of ten. In specific, we used the step-up light curve presented in the top panel of Figure \ref{fig:stepup} and we extracted the spectrum computed by our \tpho model at $t=25\ \rm ks$ after the jump. As for the initial conditions, we assumed the SED of Mrk~509 and a warm absorber with density $\nh=10^{5}\ \rm cm^{-3}$, column density $\NH=5\times10^{21}\ \rm cm^{-3}$, ionisation parameter $\log \xi = 1$, and outflow velocity $v_{\rm out} = -300\ \rm \kms$. We simulated the signal-to-noise ratio expected for a 50~ks observation taken with \xmm/RGS. The synthetic spectrum contains the time-dependent effects and it is shown in the upper left panel of Figure \ref{fig:simu}. Following the common photoionisation modelling procedure, we fit the absorption features of the warm absorber adding a \pion component and setting the $\NH$, $\log \xi$ and $v_{\rm out}$ as free parameters. The best fit, shown with a solid red line, favours a warm absorber component with a higher ionisation parameter of 1.6. We obtained a worse statistic ($\Delta C{\rm \text{-}stat}=60$) and significant residuals around the iron UTA, as shown in the left bottom panel. To correct for these residuals we added a second \pion component (best fit in green). The fit statistic improved by $\Delta C{\rm \text{-}stat}=22$ for three free parameters and two absorbers with $\log \xi$ of 1.3 and 2.0 and similar column density are found. It is thus important to include the time-dependent effects in the photoionisation modelling of AGN outflow.

For modelling extremely variable sources, time-independent photoionisation models may not be sufficiently accurate. In some cases, they might erroneously indicate the presence of a larger number of absorber components with respect to a time-dependent photoionisation analysis. As we have shown in Section \ref{sec:tpho}, the ionic concentrations follow different time evolutions based on the density and ionisation state of the gas. Thus, during specific epochs (e.g. after a strong variation of the intrinsic luminosity), it is not possible to reproduce the spectrum of a gas in non-equilibrium with a standard photoionisation modelling and the best-fitting would show residuals around some spectral features. These discrepancies will become more significant with the advent of the new generation of high-resolution X-ray instruments.

We repeat the same analysis using the up-to-date response matrix of the mission concept \arcus which is expected to reach a resolving power of R=3800 in the waveband $10-60$~\ang and an effective area of $325\; \rm cm^2$ at 19~\ang. To simulate the response of the instrument, we followed the detailed guide\footnote{\url{http://www.arcusxray.org/responses/OverallGuide.pdf}} written by the \arcus team. Due to its large effective area ($325\;\rm cm^{2}$ at 19~\ang) and resolution power in the soft X-ray band ($R=3800$), \arcus would represents a powerful instrument to study how AGN outflows respond to any luminosity variation. It will revolutionise current photoionisation modelling approaches as \chandra and \xmm did in the early 2000s.

In the top right panel of Figure \ref{fig:simu}, we show the \arcus synthetic spectrum computed using the same set-up and exposure time described above. The plot of the residual shows a strong discrepancy in the Fe UTA band which neither a single nor a multi steady-state component can reproduce. By not including the time-dependent effects in the modelling, the $C$-statistics of the best fit decrease by $\Delta C{\rm \text{-}stat}\sim1600$. In presence of a variable photoionising source it will be crucial to include the time-dependent effects in the modelling of the absorbing plasma.

The X-ray Imaging and Spectroscopy Mission \citep[{\it XRISM};][]{Tashiro18} represents the next high-resolution X-ray telescope to be launched in 2023. The soft band of the synthetic spectrum obtained with the X-ray micro-calorimeter \textit{Resolve} \citep{Ishisaki18} is shown in \ref{fig:xrism}. Similarly to the previous cases, \pion cannot characterise correctly the absorption features linked to an ionised outflow which is in a non-equilibrium photoionisation state. The statistic of the best fit decreases by $\Delta C{\rm \text{-}stat}\sim900$ with respect to the best fit obtained with \tpho. In this case, adding a second \pion component significantly improves the overall best fit ($\Delta C{\rm \text{-}stat}\sim170$) as the residual highlights in the bottom panel of Figure \ref{fig:xrism}.

\section{Summary}
\label{sec:summary}
Density represents a crucial physical quantity in modelling of photoionised plasmas and ascertaining the energetics of AGN outflows. In general, standard photoionisation models that assume a gas in constant ionisation equilibrium do not take into account time-dependent effects in the plasma. In the present work we have shown the impact of the density on the ionisation state of a gas photoionised by a variable source. Each ion has a specific recombination and photoionisation timescales which depend on both the ionisation state of the plasma and its density. When the gas is not in ionisation balance, i.e. when the $\teq\lesssim\tvar$, the shape of the transmitted spectrum is significantly affected by the time-dependent photoionisation effects.

We developed a new time-dependent photoionisation model, \tpho, which has already been implemented and publicly released with \spex v. 3.07. The model reads as input the light curve and the ionising SED and computes the time evolution of the ionic concentrations and transmitted spectrum of a specific plasma ($\nh$, $\log \xi$, $\NH$, $v_{\rm out}$, $v_{\rm turb}$). The primary goals of \tpho are:
\begin{itemize}
\item {\it to characterise the time-dependent effects}. We demonstrated that in presence of a variable ionising source if the time dependent effects are taken into account the warm absorber spectral features can be explained by a single components instead of multiple steady-state phases. Implementing the time-dependent effects in the photoionisation modelling of AGN outflow is essential with the advent of new X-ray missions (e.g., {\it XRISM} and \arcus).
\item {\it to constrain the absorber density and location}. Knowing the distance of the different absorbing clouds from the ionising source enables us to map the AGN outflows and to determine their energetics. Applying the \tpho model to a sample of bright Seyfert I galaxies which show the presence of outflows can provide important insights on the AGN feedback driven by the outflows and possibly understand the relation between ultra-fast outflows, warm absorbers and molecular outflows.
\end{itemize}
Future X-ray telescopes will allow accurate time-dependent photoionisation modelling of plasma in the presence of variable ionising sources. 
In particular, low-density, ionised outflows of variable AGN will require a model that is able to characterise the time-effects. Long monitoring campaigns of bright Seyfert I galaxies will not only help to map the outflows along the line of sight but also to better understand their launch mechanism and to estimate the mass and energy that they carry.


\begin{acknowledgments}
We thank the referee for their useful suggestions and comments. DR is grateful to Peter Kosec, Dave Huenemoerder, and Claude Canizares for insightful discussion. DR also thank Moritz Guenther for his support on the Arcus simulation. DR is supported by NASA through the Smithsonian Astrophysical Observatory (SAO) contract SV3-73016 to MIT for Support of the Chandra X-Ray Center (CXC) and Science Instruments. The CXC is operated by the Smithsonian Astrophysical Observatory for and on behalf of NASA under contract NAS8-03060. The research leading to these results has received funding from the European Union’s Horizon 2020 Programme under the AHEAD2020 project (grant agreement n. 871158). EK acknowledges XRISM Participating Scientist Program for support under NASA grant 80NSSC20K0733.
\end{acknowledgments}

%

\vspace{5mm}


\software{The \tpho model has been implemented in the X-ray fitting program \spex \citep{Kaastra22} v 3.07 (\url{https://doi.org/10.5281/zenodo.6948884}).}




\bibliography{biblio.bib}



\end{document}